\documentclass[sigconf]{acmart}
\copyrightyear{2022}
\acmYear{2022}
\setcopyright{acmlicensed}\acmConference[WWW '22]{Proceedings of the ACM Web Conference 2022}{April 25--29, 2022}{Virtual Event, Lyon, France}
\acmBooktitle{Proceedings of the ACM Web Conference 2022 (WWW '22), April 25--29, 2022, Virtual Event, Lyon, France}
\acmPrice{15.00}
\acmDOI{10.1145/3485447.3512210}
\acmISBN{978-1-4503-9096-5/22/04}

\usepackage{multirow}
\usepackage{pifont} %
\usepackage{subcaption}
\usepackage{nicefrac} %
\usepackage{cleveref} %
\usepackage{stfloats}
\usepackage{enumitem}
\usepackage{bm}
\let\vec\bm
\usepackage[english,linesnumbered,vlined,ruled,nokwfunc]{algorithm2e}
\DontPrintSemicolon
\SetKwComment{note}{$\triangleright$ }{}
\SetFuncSty{textsc}
\SetCommentSty{textit}
\SetDataSty{texttt}
\SetKwInput{Initial}{Initial}
\SetKwInput{Parameter}{Param.}
\SetKwInput{Input}{Input}
\SetKwInput{Data}{Data}
\SetKwInput{Output}{Output}
\ResetInOut{Result} 
\let\oldnl\nl%
\newcommand{\nonl}{\renewcommand{\nl}{\let\nl\oldnl}}%
\SetKw{KwAnd}{and} %
\SetKw{KwOr}{or}
\SetKw{KwXor}{xor}
\SetKw{KwNot}{not}
\SetKw{Parallel}{parallel}

\newcommand{\tg}{\mathcal{G}}
\newcommand{\tge}{\mathcal{E}}
\newcommand{\pluseq}{\mathrel{+}=}
\newcommand{\cNP}[0]{\ensuremath{\mathsf{NP}}}
\newcommand{\norm}[1]{\left\lVert#1\right\rVert}

\newcommand{\WalkGstatic}[0]{\ensuremath{\mathcal{W}}}
\newcommand{\WWalkInStatic}[0]{\ensuremath{W_{in}}}
\newcommand{\WWalkOutStatic}[0]{\ensuremath{W_{out}}}
\newcommand{\WalkG}[0]{\ensuremath{\bm{\WalkGstatic}}}
\newcommand{\WalkIn}[0]{\ensuremath{\WalkG_{in}}}
\newcommand{\WalkOut}[0]{\ensuremath{\WalkG_{out}}}
\newcommand{\WWalkIn}[0]{\ensuremath{\bm{W}_{in}}}
\newcommand{\WWalkOut}[0]{\ensuremath{\bm{W}_{out}}}

\usepackage{tikz}
\usetikzlibrary{arrows,decorations.pathmorphing}
\tikzset{
    vertex/.style={circle,draw,inner sep=01pt,minimum size=1.3em},
    vertex2/.style={circle,draw,inner sep=01pt,minimum size=1.5em},	
    vertex2g/.style={color=gray,circle,draw,inner sep=01pt,minimum size=1.5em},		
    vertexb/.style={draw,inner sep=01pt,minimum size=1.3em},	
    dedge/.style={-,> = latex', font=\footnotesize},
}
\usetikzlibrary{arrows.meta}
\begin{document}

\title{Temporal Walk Centrality: Ranking Nodes in Evolving Networks}

\author{Lutz Oettershagen}
\email{lutz.oettershagen@cs.uni-bonn.de}
 \orcid{1234-5678-9012}
\affiliation{%
  \institution{University of Bonn} %
  \city{Bonn}
  \country{Germany}
}

\author{Petra Mutzel}
\email{petra.mutzel@cs.uni-bonn.de}
 \orcid{0000-0002-2526-8762}
\affiliation{%
  \institution{University of Bonn} %
  \city{Bonn}
  \country{Germany}
}

\author{Nils M.~Kriege}
\email{nils.kriege@univie.ac.at}
 \orcid{0000-0003-2645-947X}
\affiliation{%
  \institution{University of Vienna} %
  \city{Vienna}
  \country{Austria}
}

\begin{abstract}
We propose the \emph{Temporal Walk Centrality}, which quantifies the importance of a node by measuring its ability to obtain and distribute information in a temporal network.
In contrast to the widely-used betweenness centrality, we assume that information does not necessarily spread on shortest paths but on temporal random walks that satisfy the time constraints of the network. 
We show that temporal walk centrality can identify nodes playing central roles in dissemination processes that might not be detected by related betweenness concepts and other common static and temporal centrality measures.
We propose exact and approximation algorithms with different running times depending on the properties of the temporal network and parameters of our new centrality measure. 
A technical contribution is a general approach to lift existing algebraic methods for counting walks in static networks to temporal networks.
Our experiments on real-world temporal networks show the efficiency and accuracy of our algorithms. 
Finally, we demonstrate that the rankings by temporal walk centrality often differ significantly from those of other state-of-the-art temporal centralities.
\end{abstract}




\maketitle

\section{Introduction}
Measuring the centrality of nodes in a network is a cornerstone of network analysis.
The goal is to determine the importance of nodes in the network and find the most central ones.
Various concepts of centrality have been proposed~\cite{Newman2010,Saxena2020}, and their informative value must be assessed based on a research question. 
A prime example is the \emph{PageRank} algorithm~\cite{Page1999} for ranking web pages in search engine results by performing random walks among them.
Random walks also form the basis of the classical \emph{Katz centrality}~\cite{katz1953new}, which measures node importance in terms of the number of random walks starting (or arriving) at a node, down-weighted by their length. Thereby, the Katz centrality measures the ability of a node to send out or receive information.
A different widely-used notion of centrality is \emph{betweenness}.
Freeman~\cite{Freeman1977} defines the betweenness of a node as the fraction of shortest paths between pairs of nodes that pass through it. Therefore, betweenness determines the importance of a node by its ability to pass on information.
However, considering only the shortest paths can be too restrictive since information or diseases do not necessarily spread along shortest paths.
Therefore, a betweenness centrality based on random walks has been considered~\cite{Newman2005}, which takes all walks that visit a node into account.
These centrality measures, as well as many more, are primarily designed for static networks.
However, real-world networks are often dynamic, and their topology changes over time~\cite{holme2015modern}. 
Recently, the study of centrality in \emph{temporal networks} in which temporal edges only exist at specific points in time has gained increasing attention \cite{DBLP:conf/pkdd/RozenshteinG16, BussMNR20,katztg,oettershagen2020efficient,oettershagen2022computing}. 
Important examples of such temporal networks include communication and web-based social networks such as email correspondences or human contacts~\cite{holme2015modern}. 
\\
\noindent\textbf{Our work:}
We introduce a new centrality measure for temporal networks called \emph{Temporal Walk Centrality}, which assesses the importance of a node by its ability to obtain and distribute information. 
Our new centrality measure generalizes the temporal Katz centrality as well as the degree centrality~\cite{grindrod2011communicability,katztg}.
Like the static random walk betweenness, the temporal walk centrality counts random walks passing through a node. 
However, the temporal nature of sequential events modeled by temporal networks implies causality by the forward flow of time, which needs to be respected in temporal network analysis.
For example, if there is a contact between individual $A$ and $B$ at time $t_1$, and $B$ and $C$ at time $t_2>t_1$, information may pass from $A$ to $C$ via $B$ but not vice versa.
To take this into account, we use the concept of \emph{temporal walks} that respect the flow of time in contrast to \emph{static} walks. 
Temporal walks only allow following an edge at any node if the edge exists at a time point not earlier than the arrival time at the node.
Analogously to~\cite{BussMNR20}, we distinguish between \emph{strict} and \emph{non-strict} temporal walks.
We assume that a global transition time is required to traverse edges, which is allowed to be zero.
In this case, \emph{non-strict} temporal walks can include consecutive edges with equal time stamps and may contain cycles.
If the transition time is non-zero, each edge can be contained at most once in a \emph{strict} temporal walk.
While some previously proposed temporal walk-based techniques only support strict temporal walks~\cite{DBLP:conf/pkdd/RozenshteinG16,katztg}, others are designed to take non-strict temporal walks of unbounded length into account~\cite{grindrod2011communicability}.
The temporal walk centrality is general and supports both models. We propose corresponding algorithms with different properties.

\noindent\textbf{Contributions:} We make the following contributions:
\begin{enumerate}[noitemsep,topsep=0pt] %
    \item We introduce the temporal walk centrality, which captures the ability of nodes in temporal networks to obtain and distribute information. We demonstrate that nodes with high temporal walk centrality are key players in the dissemination of information. 
    \item We present efficient algorithms for computing the temporal walk centrality with non-strict and strict temporal walks. For the first case, we propose to derive a static directed line graph from which temporal walks can be counted by general algebraic techniques. This yields an exact and iterative approximate algorithm with running time bounded by $\mathcal{O}(km^2)$, where $m$ is the number of temporal edges and $k$ the number of iterations.  For strict temporal walks, we introduce a highly efficient and scalable streaming-based algorithm with running time in $\mathcal{O}(m\cdot \tau_{max})$, where $\tau_{max}$ is the maximal number of arrival or starting times at any node.
    \item In our evaluation, we show that our approximation is efficient and achieves high-quality solutions. Furthermore, our streaming algorithm is fast even for temporal graphs with hundreds of millions of edges. Finally, our experiments show that the temporal walk centrality differs from other state-of-the-art temporal centrality measures recommending its application in scenarios where information spreads along random walks.
\end{enumerate}
Due to the space restrictions, all proofs can be found in \Cref{sec:proofs}.
\\\\
\noindent
\textbf{Relevance to Web Research:} 
Identifying and ranking nodes of
web-based social networks and communication networks according to their importance in the dissemination of information are critical and challenging tasks, especially if the considered networks are non-static and of temporal nature. Our work provides a new centrality measure that can rank nodes according to their importance in information spreading. Possible applications are the surveillance of spreading fake news or infectious diseases in temporal networks.

\section{Related Work}
There are excellent introductions to temporal graphs that include surveys on different temporal centrality measures, see, e.g., \cite{holme2015modern,santoro2011time,streamgraphs}. 
Further overviews of centralities and their applications are provided, e.g., in~\cite{white2003algorithms,das2018study,landherr2010critical,rodrigues2019network,Saxena2020,ronqui2015analyzing,gomez2019centrality}. 
In a recent work, Buß et al.~\cite{BussMNR20} discuss several variants of temporal betweenness based on shortest paths and study their theoretical complexity and practical hardness.
They consider different criteria of path lengths, such as arrival time and total travel time. They show that the running time for computing temporal betweenness using temporal shortest paths is in $\mathcal{O}(n^3\cdot T^2)$ with $n$ the number of nodes and $T$ the total number of time steps in the temporal graph. This complexity holds for strict and non-strict temporal paths.
In~\cite{tsalouchidou2019temporal}, the authors extend Brandes' algorithm~\cite{brandes2001faster} for distributed computation of betweenness centrality in temporal graphs. They introduce \emph{shortest-fastest} paths as a combination of the conventional distance and shortest duration.
The authors of~\cite{katztg,grindrod2011communicability} adapt the walk-based Katz centrality~\cite{katz1953new} to temporal graphs.
Rozenshtein and Gionis~\cite{DBLP:conf/pkdd/RozenshteinG16} incorporate the temporal character in the definition of the PageRank and consider a walk-based perspective. Hence, they obtain a temporal PageRank by replacing walks with temporal walks. 
Several papers compare temporal distance metrics as well as temporal versions of centrality measures to their static counterparts, e.g., Nicosia et al.~\cite{nicosia2013graph}, Kim and Anderson~\cite{kim2012temporal}, Tang et al.~\cite{tang2013applications,tang2010analysing}, and show that temporal approaches have advantages over static approaches on the aggregated graphs.
In~\cite{crescenzi2020finding} and \cite{oettershagen2020efficient}, the authors introduce top-$k$ algorithms for temporal closeness variants.
The authors of~\cite{haddadan2021repbublik} introduce a closeness variant based on bounded random-walks.
Related to the considered node importance is the concept of \emph{influence} in social networks, which has been studied extensively, see, e.g.,~\cite{Kempe2003,Wu2020} and references therein. Here, one is interested in a subset of the nodes that, when activated (e.g., convinced to adopt a product), have the strongest effect on the network according to some diffusion model. Finding such a set is typically \cNP-hard but can often be approximated with guarantees~\cite{Kempe2003}. Recently, dynamic graph algorithms and approaches for temporal networks have been proposed~\cite{Wu2020}.

\section{Preliminaries}
An \emph{undirected} (static) \emph{graph} $G=(V, E)$ consists of a finite set of nodes $V$ and a finite set $E\subseteq\{\{u,v\}\subseteq V\mid u\neq v\}$ of undirected edges. In a \emph{directed} (static) graph  $E\subseteq \{(u,v)\in V\times V \mid u\neq v\}$.
We use $V(G)$ to denote the set of nodes of $G$. 
The out-degree $d^+(v)$ and in-degree $d^-(v)$ of a node $v$ in a directed graph is the number of outgoing edges $(v,\cdot)$ and incoming edges $(\cdot, v)$ in $E$, respectively. %
A (static) \emph{walk} in a graph $G$ is an alternating sequence of nodes and edges connecting consecutive nodes. 
For notational convenience we sometimes omit edges. 
The length of a walk is the number of edges it contains.

A \emph{temporal graph} $\tg=(V, \tge, \delta)$ consists of a finite set of nodes $V$, a finite set $\tge$ of directed \emph{temporal edges} $e=(u,v,t)$ with $u$, $v$ in $V$, $u\neq v$, and \emph{availability time} (or \emph{time stamp}) $t \in \mathbb{N}$. 
The parameter $\delta\in \mathbb{N}$ is a global transition delay of the edges that defines how long it takes to transmit information via an edge.   
The starting time of an edge $e=(u,v,t)$ at node $u$ is $t$, and the arrival time at node $v$ is $t+\delta$. 
The number of edges is not necessarily polynomially bounded by the number of nodes because each pair of nodes can be connected at several points in time. 
For $v\in V$ let $T(v)$ be the set of availability times of edges incident to $v$.
We define $T(\tg)= \{t, t+\delta\mid (u,v,t)\in\tge\}$.
Furthermore, let $\tau^-_{max}$ ($\tau^+_{max}$) denote the maximum number of distinct arrival (starting) times at any of the nodes.
The in- and out-degree of a node in a temporal graph is the total numbers of incoming and outgoing edges over all time steps.
We only consider directed temporal graphs and model undirectedness using a forward- and a backward-directed edge with equal time stamps for each undirected edge.

A \emph{temporal walk} $\omega$ in a temporal graph $\tg$ is an alternating sequence $(v_1, e_1,\ldots,e_\ell,v_{\ell+1})$ of nodes and temporal edges connecting consecutive nodes, 
where $e_i=(v_i,v_{i+1},t_i)\in \tge$ and $t_i+\delta\leq t_{i+1}$ for all $i \in \{1,\dots,\ell\}$. 
We may omit nodes for notational convenience. 
Depending on $\delta$, we distinguish \emph{strict} and \emph{non-strict} temporal walks, where for $\delta=0$, the temporal walks are non-strict.
We denote the length of a temporal walk $\omega$ by $|\omega|=\ell$. 
For $\delta>0$, the length of a (strict) temporal walk is bounded by $T(\tg)$, where for $\delta=0$, there is no general upper bound on the length of non-strict temporal walks.
It is common to restrict temporal walks to a time interval $[a,b]$ with $a,b\in\mathbb{N}$ and $a\leq b$.
This case is covered by running our algorithms on the temporal subgraph containing only the edges $(u,v,t)\in\tge$ for which $a\leq t$ and $t+\delta\leq b$.
Moreover, our definitions and algorithms can be easily extended to the case of individual transition times $\delta_e$ for each edge $e$ instead of the global parameter $\delta$.
\Cref{table:notation} in \Cref{appendix:notation} summarizes our notation.
\section{Temporal Walk Centrality} %

The intuition of our new centrality measure is that nodes are regarded as important if they are involved in the process of passing information. 
The time stamps of the temporal edges imply causality and direct the information flow in the network. Therefore, we measure the contribution of a node by means of temporal walks respecting such aspects.
We define the centrality of a node $v$ as the number of temporal walks passing through $v$, where the temporal walks are weighted depending on their length and temporal structure.
We formalize this concept before proposing our new centrality measure, and define a weight function for temporal walks similar to \cite{katztg} and \cite{nguyen2018continuous}.
\begin{definition}[Temporal walk weight]\label{def:tmp_walk_weight}
    Let $\tg=(V,\tge,\delta)$ be a temporal graph, and $\omega=(e_1,\ldots,e_\ell)$ a temporal walk in $\tg$.
    We define the weight of a temporal walk $\omega$ as 
    \begin{equation}
        \tau_\Phi(\omega) = \prod_{i=1}^{\ell-1}\Phi(t_i+\delta, t_{i+1}),
    \end{equation}
where the function $\Phi\colon\mathbb{N}\times \mathbb{N}\rightarrow \mathbb{R}$ is a time depended weight function.
We define $\tau_\Phi(\omega')=1$ for walks $\omega'$ of length zero and one. 
\end{definition}
We discuss concrete examples of the function $\Phi$ later in this section.
First, we define the total weight of incoming and outgoing temporal walks at each node $v\in V$.
\begin{definition}
    Let $\tg=(V,\tge, \delta)$ be a temporal graph, and let $\WalkIn(v,t)$ and $\WalkOut(v,t)$ be the sets of incoming and outgoing temporal walks, resp., at node $v$ and time $t$. 
    We define
    \begin{equation*}
         \WWalkIn(v,t) = \hspace{-1.2em}\sum_{\omega\in \WalkIn(v,t)}\hspace{-1.2em}\tau_{\Phi_{in}}(\omega) 
         \hspace{1em}\text{and}\hspace{1em} 
         \WWalkOut(v,t) = \hspace{-1.2em}\sum_{\omega\in \WalkOut(v,t)}\hspace{-1.2em}\tau_{\Phi_{out}}(\omega).
    \end{equation*}
\end{definition}
The temporal walk weight functions $\tau_{\Phi_{in}}$ and $\tau_{\Phi_{out}}$ allow to weight incoming and outgoing walks independently.
Using $\WWalkIn$ and $\WWalkOut$, we now define the temporal walk centrality.

\begin{definition}[Temporal Walk Centrality]\label{def:twc}
    Let $\tg=(V,\tge)$ be a temporal graph. %
    We call
    \begin{equation*}
    C(v) = \sum_{t_1,t_2\in{T(\tg)}, t_1\leq t_2} \left(\WWalkIn(v,t_1) \cdot \WWalkOut(v,t_2)\cdot \Phi_m(t_1, t_2)\right)
    \end{equation*}
    the \emph{temporal walk centrality} of node $v\in V$.
\end{definition}
The time-depended weight function $\Phi_m$ is, similarly to $\Phi_{in}$ and $\Phi_{out}$, used to weight the time between obtaining and distributing information at node $v$.
Using these functions, we can weight temporal walks depending on different structural and temporal properties.
We propose the following weight functions: %
\begin{enumerate}
    \item \emph{Weighting based on length:} By setting $\Phi(t_1,t_2)=\alpha$, with $0<\alpha<1$, for all $t_1,t_2\in \mathbb{N}$, we obtain an exponential decay in the length of the walk, i.e., $\tau_\Phi(\omega) = \alpha^{|\omega|-1}$. In this case, we set $\Phi_m(t_1,t_2)=1$.
    Hence, long walks are down-weighted compared to short walks controlled by the parameter $\alpha$.
    In the following, we denote this variant with $\Phi_\alpha$.
    \item \emph{Weighting based on waiting time:} The value of information decreases with time, and we might want to weight the ability to quickly distribute new information high. In this case, we define
    $\Phi(t_1,t_2)=\frac{1}{1+t_2-t_1}$. The weight decreases with increasing waiting time at every node and remains stable at nodes, where information is passed through without delay.
    In the following, we denote this variant with $\Phi_t$. 
\end{enumerate}
In both examples, we set $\Phi_{in}(t_1,t_2)=\Phi_{out}(t_1,t_2)=\Phi(t_1,t_2)$.
Notice, that our definition is general and supports further weight functions such as a combination of (1) and (2), where we set  $\Phi(t_1,t_2)=\frac{\alpha}{1+t_2-t_1}$ for all $t_1,t_2\in \mathbb{N}$. Another example would be to use different values for $\alpha$ for incoming and outgoing walks.
Finally, we can achieve an exponential decay in the total waiting time or duration by choosing a suitable time-dependent exponential function for $\Phi$.

To see how the TemporalWalkCentrality generalizes the temporal Katz and degree centrality, consider the following.
If we fix $\WWalkIn(v,t)=1$, and use the weighting function $\Phi(a,b)=\alpha$ it follows that $C(v)$ equals the temporal Katz centrality. If we, additionally,  only consider walks of length one, $C(v)$ equals the outdegree centrality. Note that a walk length restriction can be added straightforwardly. 
\subsection{Comparison to Other Centrality Measures}\label{sec:comparison}
We compare our new temporal walk centrality to state-of-the-art centrality measures for temporal and static graphs
using an example graph and a subgraph of a real-world communication network.

\paragraph{First example:}
Consider the temporal graph $\tg=(V,\tge,\delta)$ with $\delta=1$ and availability times shown in \Cref{fig:example1}. 
Nodes $a$, $f$, and $g$ cannot pass any information and are marked in gray. 
Node $a$ does not have any incoming edges. Thus, it cannot pass any information. Similarly, node $g$ does not have any outgoing edges.
For node $f$ the only outgoing edge has availability time two. However, information can reach $f$ only via edge $(e,f,5)$ at time 6 at the earliest.
Hence, node $f$ cannot pass on any information in a dissemination process.

\Cref{table:comparision} shows a comparison of the resulting rankings of the nodes of the temporal graph shown in \Cref{fig:example1}.
The rankings are computed with different temporal and static centrality measures.
Temporal betweenness is the shortest paths variant from \cite{BussMNR20}. It counts the number of temporal shortest paths that visit a node.
The temporal PageRank~\cite{DBLP:conf/pkdd/RozenshteinG16} is a temporal version of the static PageRank centrality~\cite{brin1998anatomy}. 
Temporal Katz centrality is defined in~\cite{katztg}. The nodes are ranked according to the number of incoming temporal walks weighted by a time depended exponential decay function.
The temporal closeness centrality is a harmonic closeness variant using the shortest duration as distance function~\cite{oettershagen2020efficient,oettershagen2022computing}. 
Our temporal walk centrality is computed with $\tau_{\Phi_{in}}(\omega)=\tau_{\Phi_{out}}(\omega)=1$ for all walks $\omega$ in $\tg$, and we set $\Phi_m(t,t')=1$ for all $t,t'\in \mathbb{N}$.

We observe that only the temporal walk centrality can identify the nodes $a$, $f$, and $g$ as nodes that have no capability of passing information. 
Notice that temporal betweenness assigns the nodes to only three different ranks because it only considers the temporal \emph{shortest} paths. Therefore, it does not reveal the difference between, e.g., node $d$ and nodes $a$, $f$, or $g$, although $d$ may play an essential role in a dissemination process while the other cannot. 
Similarly, the static betweenness assigns all nodes to even only two ranks. The reason is that the computation of static shortest paths ignores the temporal restrictions. 
We can see similar problems for the other centrality measures, e.g., temporal Katz and temporal closeness rank nodes $g$ and $a$, respectively, highest. Similarly, the degree, temporal PageRank, static closeness, and static random walk betweenness fail to rank the nodes according to our idea of important nodes that can distribute information efficiently.
In conclusion, only temporal walk centrality can distinguish the nodes that are important in dissemination processes. 
It ranks the nodes according to the intuition that nodes that can be reached easily and can reach other nodes are important. 

\begin{figure*}[ht]
    \captionsetup[subfigure]{aboveskip=-1pt,belowskip=-1pt}
    \centering
    \begin{subfigure}{.23\linewidth}
        \centering
        \includegraphics[width=1\linewidth]{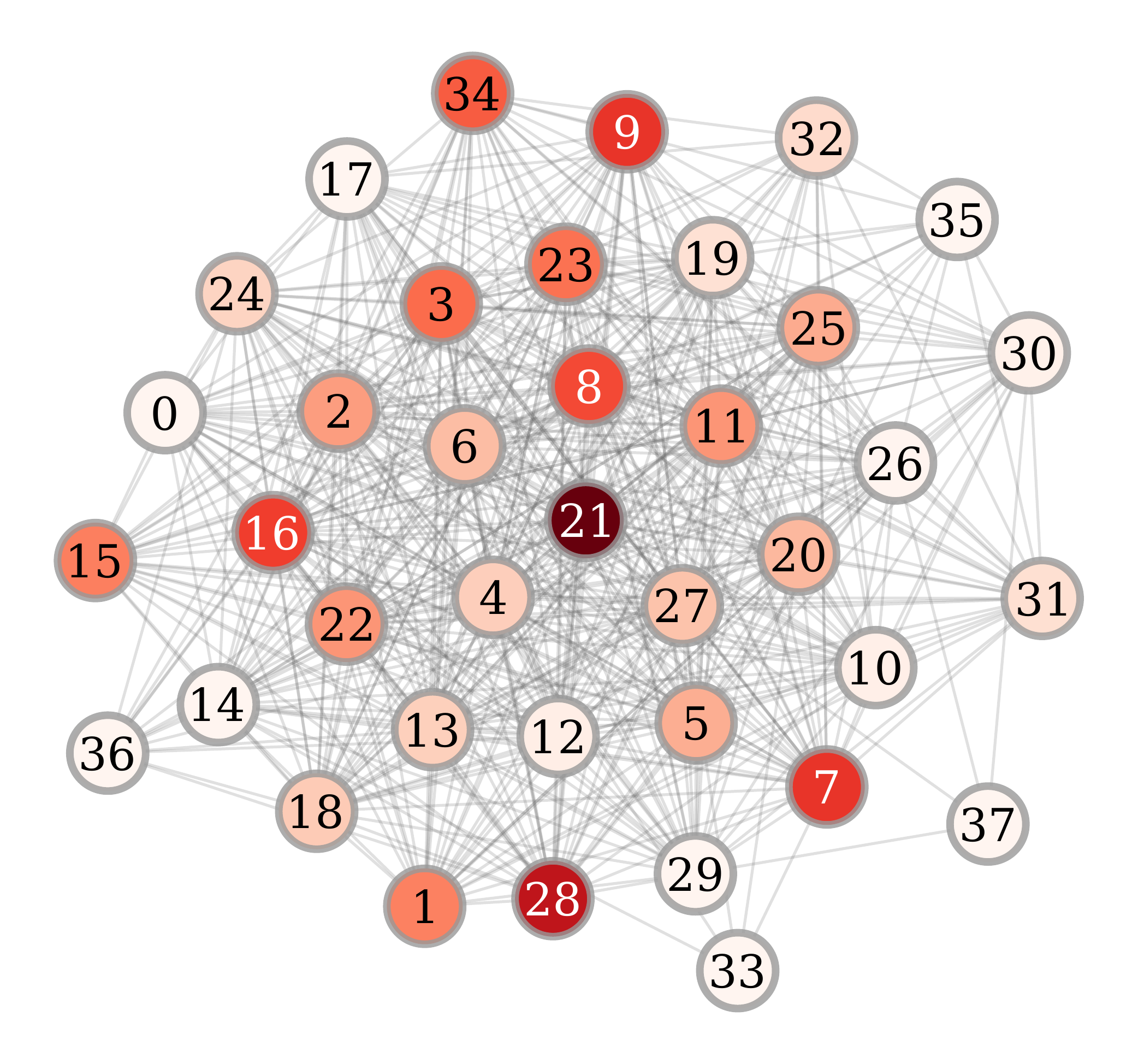}
        \caption{Temporal walk centrality. }
        \label{fig:enron_subgraphs:twc}
    \end{subfigure}\hfil%
    \begin{subfigure}{.23\linewidth}
        \centering
        \includegraphics[width=1\linewidth]{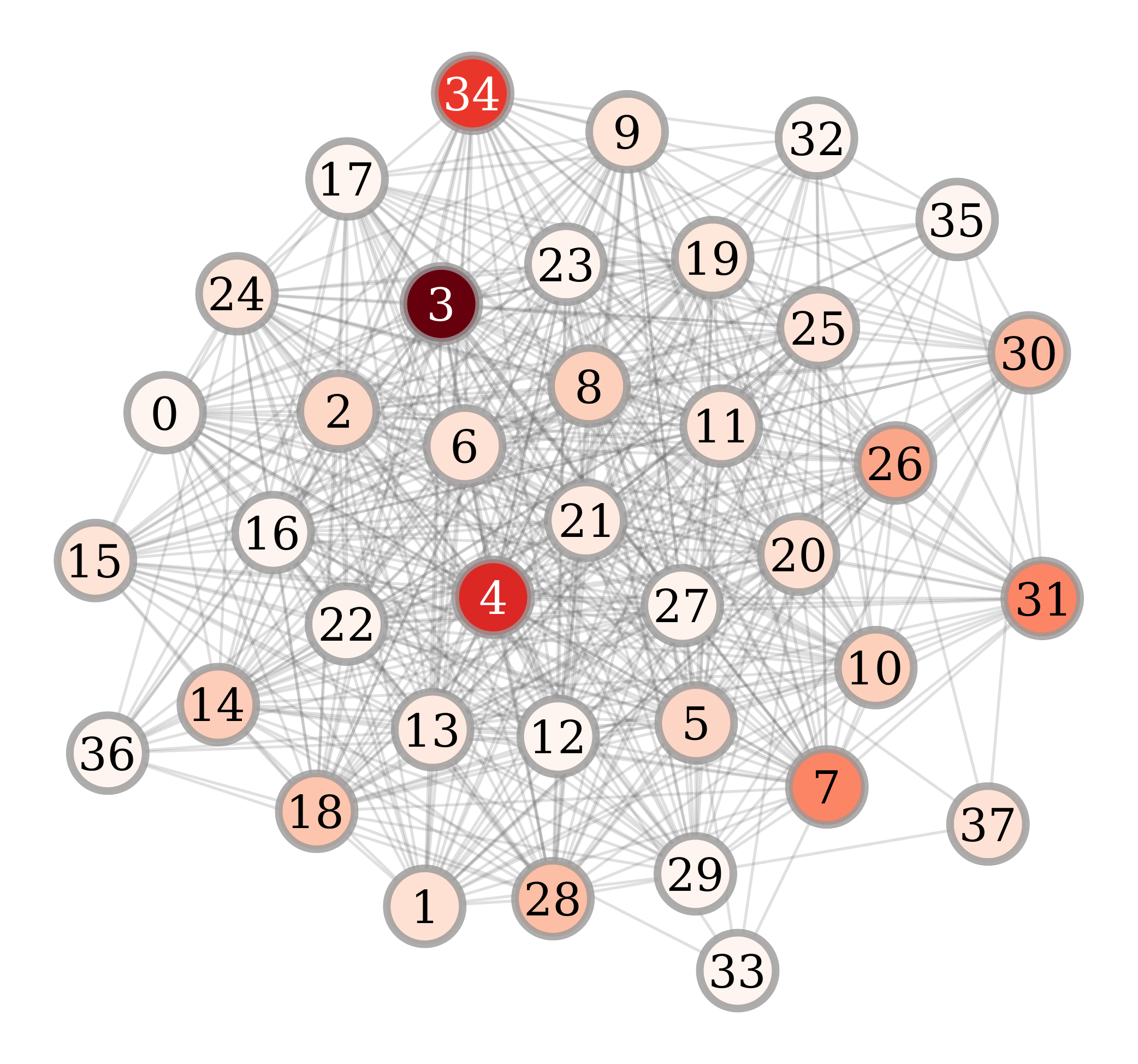}
        \caption{Temporal betweenness.}
        \label{fig:enron_subgraphs:tb}
    \end{subfigure}\hfil%
    \begin{subfigure}{.23\linewidth}
        \centering
        \includegraphics[width=1\linewidth]{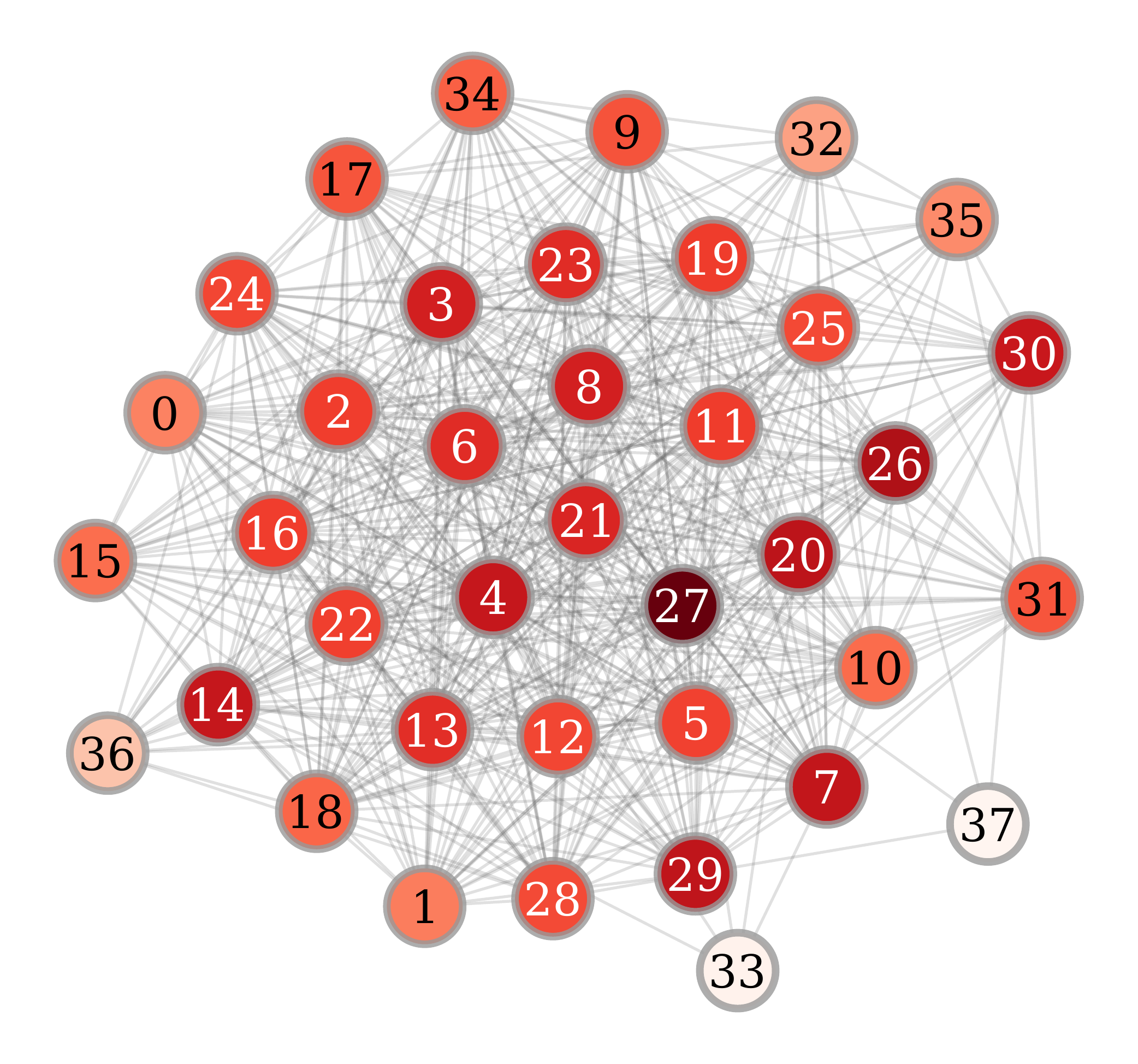}
        \caption{Static walk betweenness.}
        \label{fig:enron_subgraphs:sb}
    \end{subfigure}
    \caption{Induced subgraph of the \emph{Enron} email network consisting of 38 nodes and 541 temporal edges. The nodes are colored according to their centrality value with darker color meaning higher centrality.}
    \label{fig:enron_subgraphs}
    \Description[Another example.]{The figure shows three plots of a subgraph of an email network.}
\end{figure*}

\paragraph{Enron email network:}
We consider an induced subgraph of the \emph{Enron} email network~\cite{klimt2004enron}. The network represents the email communication in a company, where nodes are employees, and temporal edges represent emails. 
The subgraph is an ego-network with radius one of the employee represented by node zero.
\Cref{fig:enron_subgraphs} shows the network with different centrality measures, where a darker node color means higher centrality. 
\Cref{fig:enron_subgraphs:twc} illustrates the temporal walk centrality with $\Phi_\alpha$ and $\alpha=1$.
We compare it with temporal betweenness centrality (\Cref{fig:enron_subgraphs:tb}) and the static random walk betweenness (\Cref{fig:enron_subgraphs:sb}).
The former assigns high centrality values only to nodes that are part of many shortest temporal paths. 
Therefore, only a few nodes get a relatively high centrality value, namely the nodes $3$, $4$, and $34$. 
However, nodes that have a high capability to pass information, e.g., node 21, are assigned a low centrality value.
On the other hand, in \Cref{fig:enron_subgraphs:sb} we see that the static random walk centrality, which does not respect the temporal properties of the network, assigns high values to many nodes. The reason is that considering static walks instead of temporal walks leads to over-counting node visits by ignoring the restricted temporal reachability. Hence, many nodes in \Cref{fig:enron_subgraphs:sb} obtained a similar centrality value resulting in less information about the importance of the nodes.
\begin{figure}
    \centering
   \begin{tikzpicture}[scale=1.0]
    \begin{scope}[every node/.style={circle,thick,draw,minimum size=6mm,inner sep=0pt,font=\footnotesize}]
    \node[fill=lightgray] (A) at (1,1) {$a$}; 
    \node (B) at (2,0) {$b$};
    \node (C) at (3,1) {$c$};
    \node (D) at (4,0) {$d$};
    \node (E) at (5,1) {$e$};
    \node[fill=lightgray] (F) at (6,0) {$f$};
    \node[fill=lightgray] (G) at (7,1) {$g$};
    \end{scope}
    
    \begin{scope}[>={Stealth[black]},
    every node/.style={circle},
    every edge/.style={draw=black,thick}]
    \path [->] (A) edge node[below,pos=.2] {$1$} (B);
    \path [->] (B) edge node[below,pos=.8] {$3$} (C);
    \path [->] (A) edge node[above] {$2$} (C);
    \path [->] (C) edge node[below,pos=.2] {$3$} (D);
    \path [->] (C) edge node[above] {$3$} (E);
    \path [->] (D) edge node[below,pos=.8] {$4$} (E);
    \path [->] (E) edge node[below,pos=.2] {$5$} (F);
    \path [->] (F) edge node[below,pos=.8] {$2$} (G);
    \path [->] (E) edge node[above] {$5$} (G);
    \end{scope}
    \end{tikzpicture}
    \caption{Temporal graph $\tg$ with the availability times shown at the edges and $\delta=1$. 
    Nodes that cannot pass on information are marked in gray.}
    \label{fig:example1}
    \Description[A small example.]{The figure shows a small graph as example. The graph has seven nodes and nine edges. Three of the nodes cannot pass any information.}
\end{figure}
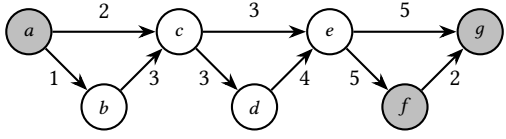
\begin{table}[t]
    \centering
    \caption{Node rankings for the temporal graph shown in \Cref{fig:example1} obtained by different centrality measures.%
    }
    \label{table:comparision}
    \resizebox{1\linewidth}{!}{\renewcommand{\arraystretch}{1.0}\setlength{\tabcolsep}{3pt}
        \begin{tabular}{lcl}\toprule
            \textbf{ Centrality }          &     & \textbf{ Node Ranking}\\ \midrule
            {Temporal Walk}                &          & \bf 1: $e$ \ 2: $c$ \ 3: $d$ \ 4: $b$ \ 5: $afg$\\ \midrule
            \multicolumn{2}{l}{Temporal Betweenness~\cite{BussMNR20}}    & \bf 1: $c$ \ 2: $e$ \  3: $abdfg$ \\
            \multicolumn{2}{l}{Temporal PageRank~\cite{DBLP:conf/pkdd/RozenshteinG16}}   & \bf1: $ce$ \  2: $a$ \  3: $fg$ \  4: $bd$\\ 
            \multicolumn{2}{l}{Temporal Katz~\cite{katztg}}             &\bf 1: $g$ \  2: $f$ \  3: $e$ \  4: $c$ \  5: $d$ \  6: $b$ \  7: $a$\\  
            \multicolumn{2}{l}{Temporal Closeness~\cite{oettershagen2020efficient}}            & \bf1: $a$ \  2: $c$ \  3: $b$ \  4: $e$ \  5: $d$ \  6: $f$ \  7: $g$\\  \midrule
            \multicolumn{2}{l}{In-Degree }                   & \bf 1: $ceg$ \  2: $bdf$ \  3: $a$ \\ 
            \multicolumn{2}{l}{Out-Degree }                  & \bf 1: $acd$ \  2: $bdf$ \  3: $g$ \\ 
            \multicolumn{2}{l}{Static Betweenness~\cite{Freeman1977}}            & \bf 1: $ce$ \  2: $abdfg$ \\
            \multicolumn{2}{l}{Static Harmonic Closeness~\cite{marchiori2000harmony}}   & \bf 1: $a$ \  2: $c$ \  3: $b$ \  4: $de$ \  5: $f$ \  6: $g$\\ 
            \multicolumn{2}{l}{Static Random Walk Betweenness~\cite{Newman2005}}& \bf 1: $ce$ \  2: $d$ \  3: $bf$ \  4: $ag$\\ 
            \bottomrule
        \end{tabular}
    }
\end{table}

\section{Computation of the Temporal Walk Centrality}\label{sec:algorithms}
Computing the walk centrality of a node $v\in V$ involves counting the weighted in- and outgoing walks at $v$ over time.
In \Cref{def:twc}, $\WWalkIn$ can be interpreted as a matrix that contains the weighted sum of the walks that arrive at node $v$ at time $t$. Analogously, we have the matrix $\WWalkOut$ for the outgoing walks.
In the following, we first describe several methods for computing these matrices and then how to calculate the walk centrality from them. 

\Cref{tab:overview} gives an overview of the different algorithms, their running time, and properties.
Notice that our algorithms perform on par or favorable compared to related state-of-the-art algorithms.
The algorithm proposed in~\cite{Newman2005} for the static random walk betweenness has a running time in  $\mathcal{O}((|E|+|V|)\cdot |V|^2)$ and space complexity in $\mathcal{O}(|V|^2)$ for a static graph $G=(V,E)$. 
Furthermore, the algorithm proposed in~\cite{BussMNR20} for computing temporal betweenness using temporal shortest paths has a running time in $\mathcal{O}(|V|^3\cdot T^2)$ with $T$ the total number of time steps in a temporal graph $\tg=(V,\tge,\delta)$, and a space complexity of $\mathcal{O}(|V|\cdot T + |\tge|)$.

For the analysis of our algorithms for computing the temporal walk centrality, we assume $|\tge|\geq \nicefrac{1}{2}|V|$, which holds unless isolated nodes exist. Since an isolated node $v$ is not involved in non-trivial walks and has $C(v)=0$, we can safely delete all isolated nodes in a preprocessing step.

\newcommand{\cmark}{{\ding{51}}}%
\newcommand{\xmark}{{\ding{55}}}%
\begin{table}[t]
    \centering
    \caption{Overview of algorithms for computing the temporal walk centrality and their properties. Here, 
        $\gamma < 2.373$ is the exponent of matrix multiplication, $k$ the number of fixed-point iterations, $e=|E(\mathit{DL}(\mathbf{G}))| \leq |\tge|^2$ the number of edges in the directed line graph, and
        $\tau_{max}$ the largest cardinality of availability or arrival times at a node.
    }
    \label{tab:overview}\renewcommand{\arraystretch}{1.1}\setlength{\tabcolsep}{5pt}
    \resizebox{\linewidth}{!}{\setlength{\tabcolsep}{3pt}
        \begin{tabular}{llcccc}\toprule
            \textbf{ Method } & \textbf{ Sec.}         & \textbf{Running Time} & \textbf{Space}  & \textbf{Non-strict} & \textbf{Exact} \\ \midrule
            \textsc{DlgMa} & \ref{sec:dlg_inv}  & $\mathcal{O}(|\tge|^\gamma)$  & $\mathcal{O}(|\tge|^2)$  & \cmark & \cmark\\
            \textsc{Approx}& \ref{sec:dlg_fpi}  & $\mathcal{O}(k(|\tge|+e))$  &  $\mathcal{O}(|\tge|+e)$  & \cmark & \xmark\\
            \textsc{Stream}& \ref{sec:streaming}   & $\mathcal{O}(|\tge|\cdot \tau_{max})$ & $\mathcal{O}(|V|\cdot \tau_{max})$ & \xmark & \cmark\\
            \bottomrule
        \end{tabular}
    }
    \Description{Overview over the properties of the algorithms.}
\end{table}

\subsection{Directed Line Graph Expansion}\label{sec:dlg}
Counting (weighted) walks in static networks can conveniently be realized in terms of basic linear algebra operations. Polynomial-time computable closed-form expressions are well-known supporting walks of unbounded length when long walks are sufficiently down-weighted to guarantee convergence~\cite{katz1953new,Newman2010}.
We lift the algebraic methods for walk counting to counting temporal walks by means of the \emph{directed line graph expansion}.
Variants of directed line graph expansions have been previously used for survivability and reliability analysis~\cite{liang2016survivability,khanna2020two}, and for graph kernels~\cite{oettershagen2020temporal}. These variants support only strict temporal walks. In contrast, we allow temporal walks that can traverse the same edge multiple times when the transition time is zero leading to a potentially infinite number of temporal walks. Moreover, our definition uses directed graphs and we do not add additional start and sink vertices as in~\cite{liang2016survivability,khanna2020two}.
\begin{definition}[Directed line graph expansion]\label{def:dlg}
    Given a temporal graph $\tg=(V,\tge,\delta)$, the \emph{directed line graph expansion}
	$\mathit{DL}(\tg)=(V',E')$ is the directed graph, where  
	every temporal edge $(u,v,t)$ in $\tge$ is represented by a node $n^{t}_{uv}$, 
	and there is an edge from $n^{t}_{uv}$ to  $n^{s}_{xy}$ if $v=x$ and $t+\delta\leq s$.
\end{definition}
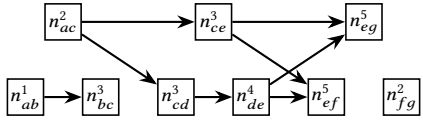
\begin{figure}[tb]
    \centering
   \begin{tikzpicture}[scale=1.0]
    \begin{scope}[every node/.style={rectangle,semithick,draw,minimum size=4.8mm,inner sep=0pt,font=\footnotesize}]
    \node (A) at (.5,1) {$n^2_{ac}$}; 
    \node (B) at (0,0) {$n^1_{ab}$};
    \node (C) at (1,0) {$n^3_{bc}$};
    \node (D) at (2,0) {$n^3_{cd}$};
    \node (E) at (2.5,1) {$n^3_{ce}$};
    \node (F) at (3,0) {$n^4_{de}$};
    \node (G) at (4,0) {$n^5_{ef}$};
    \node (H) at (4.5,1) {$n^5_{eg}$};
    \node (I) at (5,0) {$n^2_{fg}$};
    \end{scope}
    
    \begin{scope}[>={Stealth[black]},
    every node/.style={circle},
    every edge/.style={draw=black,thick}]
    \path [->] (B) edge (C);
    \path [->] (A) edge (D);
    \path [->] (A) edge (E);
    \path [->] (D) edge (F);
    \path [->] (F) edge (G);
    \path [->] (F) edge (H);
    \path [->] (E) edge (G);
    \path [->] (E) edge (H);
    \end{scope}
    \end{tikzpicture}
    \caption{Directed line graph representation of the temporal graph $\tg$ shown in \Cref{fig:example1}.}
    \label{fig:dlg}
    \Description{The figure shows an directed line graph for the temporal graph of \Cref{fig:example1}.}
\end{figure}
\Cref{fig:dlg} illustrates the concept of the directed line graph expansion.
Clearly the number of nodes of $\mathit{DL}(\tg)$ is $|\tge|$. The number of edges is maximal when at each node of the temporal graph every incoming edge can be combined with all outgoing edges, which is, for example, the case when the transition time $\delta$ is zero and all edges have the same time stamp. Then the number of edges in the directed line graph expansion of $\tg=(V,\tge,\delta)$ is 
$|E(\mathit{DL}(\tg))| = \sum_{v \in V(\tg)} d^-(v) \cdot d^+(v) = \mathcal{O}(|\tge|^2).$
This corresponds to the number of edges in the directed line graph of the underlying static graph with parallel edges~\cite{Harary1960}.

The walks in the directed line graph are closely related to the temporal walks in the original temporal graph.
We will use this relation for algebraic weighted walk counting and establish the correspondence formally.
\begin{lemma}\label{lemma:dlg_walks}
  Let $\tg$ be a temporal graph and $\ell \geq 1$. 
  Moreover, let $\WalkGstatic_\ell(G)$ be the walks of length $\ell$ in the graph $G$ 
  and $\WalkG_\ell(\tg)$ the temporal walks in the temporal graph $\tg$.
  There is a bijection $\Gamma\colon\WalkGstatic_{\ell-1}(\mathit{DL}(\tg))\to\WalkG_\ell(\tg)$ given by
 \begin{equation*}
  \left(n^{t_1}_{v_1v_2}, n^{t_2}_{v_2v_3}, \ldots, n^{t_\ell}_{v_\ell v_{\ell+1}}\right)
  \mapsto (v_1, e_1, v_2, \ldots, e_\ell, v_{\ell+1})
 \end{equation*}
 with $e_i=(v_i,v_{i+1},t_i)$ for $i \in\{1,\dots,\ell\}$. 
\end{lemma}

We identify the temporal walks starting at a specific node and time in a temporal graph with walks starting at several different nodes in its directed line graph expansion.

\begin{corollary}\label{cor:tmp_walk_to_walk}
    The temporal walks of length $\ell\geq1$ in a temporal graph $\tg$ starting (ending) at the node $v$ at time $t$ are in one-to-one correspondence with the walks of length $\ell-1$ in $\mathit{DL}(\tg)$ starting (ending) at the nodes $X_{out}(v,t)$ ($X_{in}(v,t)$), where
    \begin{align*}
    X_{out}(v,t) &= \{n^{s}_{uw} \in V(\mathit{DL}(\tg)) \mid v = u \wedge t = s\} \\
    X_{in}(v,t)  &= \{n^{s}_{uw} \in V(\mathit{DL}(\tg)) \mid v = w \wedge t = s+\delta\}.
    \end{align*}
\end{corollary}

We endow the directed line graph expansion with edge weights such that the weight of a walk corresponds to the temporal walk weight of its temporal walk according to \Cref{def:tmp_walk_weight}. 
In a static graph with edge weights $w\colon E \to \mathbb{R}$, the weight of a walk $\omega=(e_1,e_2,\dots,e_\ell)$ is $w(\omega) = \prod_{i=1}^{\ell}w(e_i)$. We annotate an edge $e=\left(n^{t}_{uv}, n^{s}_{vw}\right)$ in the directed line graph by $w_\Phi(e)=\Phi(t+\delta, s)$.
\begin{lemma}\label{lemma:dlg_weights}
 Let $\omega$ be a walk in the directed line graph expansion $\mathit{DL}(\tg)$ of the temporal graph $\tg$, then we have
 $w_\Phi(\omega)=\tau_\Phi(\Gamma(\omega))$.
\end{lemma}
The combination of these results allows to compute the temporal walk centrality.
To this end, we count the in- and outgoing walks in the directed line graph expansion weighted by $\Phi_{in}$ and $\Phi_{out}$, respectively, and apply \Cref{cor:tmp_walk_to_walk} to derive the corresponding values for the temporal graph.
More precisely, let $\WWalkOutStatic(v,\ell)$ denote the sum of weighted walks of length $\ell$ in the static graph starting at the node $v$. Then, we obtain 
\begin{equation*}
\WWalkOut(v,t) = \hspace{-1em}\sum_{x \in X_{out}(v,t)} \hspace{-1em} \WWalkOutStatic(x),\quad \WWalkOutStatic(x):=\sum_{\ell=0}^{\infty} \WWalkOutStatic(x,\ell).
\end{equation*}
The value $\WWalkIn$ can be obtained similarly by using the set $X_{in}(v,t)$ and incoming weighted walk counts $\WWalkInStatic(x)$.
Counting weighted walks in (static) graphs can be realized by means of matrix methods. However, when the transition time $\delta$ is non-zero, the directed line graph is acyclic, since an edge can only be traversed at most once by a walk. 
Efficient algorithms for this case are discussed in~\Cref{sec:streaming} and \Cref{sec:acyclic}.

\subsubsection{Computation by matrix inversion}\label{sec:dlg_inv}
Let $\vec{A}$ be the weighted adjacency matrix of a graph $G$ with weights $w$, where $a_{uv}=w(u,v)$ if $(u,v) \in E$ and $0$ otherwise.
It is well known that the entry $a^{(\ell)}_{uv}$ of $\vec{A}^\ell$ is the sum of weighted walks of length $\ell$ from node $u$ to node $v$.
Hence, we have $\WWalkOutStatic(x,\ell) = \left[\vec{A}^\ell\vec{1}\right]_x$ and $\WWalkOutStatic(x) = \left[\left(\sum_{\ell=0}^{\infty}\vec{A}^\ell\right)\vec{1}\right]_x$.
The sum of matrix powers is know as Neumann series and the identity $\sum_{\ell=0}^{\infty}\vec{A}^\ell= (\vec{I}-\vec{A})^{-1}$ holds if the sum converges. This is guaranteed when $\rho(\vec{A})<1$, where $\rho$ denotes the \emph{spectral radius}, i.e., the largest absolute value of an eigenvalue. In this case all weighted walks without length bound can be counted by inversion of an $n\times n$ matrix, where $n=|\tge|$. Exact matrix inversion is in time $\mathcal{O}(n^\gamma)$, where $\gamma < 2.373$ is theoretically possible~\cite{AlmanW21} by improving the seminal work of Coppersmith and Winograd~\cite{Coppersmith1987}. 
\begin{lemma}
 The weighted temporal walks in a temporal graph $\tg=(V,\tge,\delta)$ can be counted exactly in $\mathcal{O}(|\tge|^\gamma)$ time with space $\mathcal{O}(|\tge|^2)$, where $\gamma < 2.373$ is the exponent of matrix multiplication.
\end{lemma}
In practice, roughly cubic running time is expected. 
For large graphs this is prohibitive even when the directed line graph is sparse as the inverse of a sparse matrix not necessarily is sparse.

\begin{algorithm}[t]
    \label[algorithm]{alg:dlg:approx}
    \caption{Approximate counting weighted outgoing walks in directed graphs.}
    \Input{Directed weighted graph $G$, error-tolerance $\varepsilon$.}
    \Output{Weighted walk counts $\WWalkOutStatic(v)=[\vec{r}]_v$.}
    Initialize $\vec{A}$ from $G$ and verify $\rho(\vec{A})<1$ \;
    $\vec{v} \gets \vec{1}$; $\vec{r} \gets \vec{v}$ \note*[r]{Length $0$ walks}
    \Repeat{$\norm{\vec{v}}_1 < \varepsilon$} {
        $\vec{v} \gets \vec{A} \vec{v}$\;
        $\vec{r} \gets \vec{r}+\vec{v}$ \;
    }
\end{algorithm}

\subsubsection{Approximation by fixed-point iteration}\label{sec:dlg_fpi}
We propose an approximation algorithm based on iterated matrix-vector multiplication which benefits from algorithms and datastructures for sparse matrices.
\Cref{alg:dlg:approx} approximates $\WWalkOutStatic(x)$ and can analogously be applied to compute $\WWalkInStatic(x)$ by reversing all edges, or simply transposing the weighted adjacency matrix in a preprocessing step.
\begin{lemma}
 The weighted temporal walk counts of a temporal graph $\tg=(V,\tge,\delta)$ can be approximated in time $\mathcal{O}(k(|\tge|+e))$ and space $\mathcal{O}(|\tge|+e)$, where $e=|E(\mathit{DL}(\tg))|$ and $k$ is the number of iterations required to meet the error tolerance.
\end{lemma}
In practice, the directed line graph expansion is often sparse with $|E(\mathit{DL}(\mathbf{G}))| \ll |\tge|^2$ and we expect significant advantages of the approximate method in this case. We verify this hypothesis experimentally in \Cref{sec:experiments}.

\subsection{Streaming Algorithm}\label{sec:streaming}
In the case of $\delta>0$ the directed line graph is acyclic and we can count the weighted walks in linear time by traversing it in a bottom-up fashion, cf.~\Cref{sec:acyclic}.
However, we can avoid the construction of the directed line graph representation by directly operating on the temporal graph in \emph{edge stream representation}, where the edges are given in chronological order (ties are broken arbitrarily).
The edge stream representation has been successfully applied in earlier works for temporal paths computation~\cite{wu2014path,mutzel2019enumeration}. 
We use two passes over the edge stream.
In a forward pass, we compute the weights of incoming walks at each node and in a backward pass of outgoing walks.
\Cref{alg:streaming_algorithm} processes the edges in chronological order to compute the matrix $\WWalkIn$.
Hence, when processing a temporal edge $(u,v,t)$ the incoming walk counts for the node $u$ at all arrival times before $t$ are correctly computed since all edges $(\cdot,u,t')$ with $t'+\delta\leq t$ have already been processed.
\begin{algorithm}
    \label[algorithm]{alg:streaming_algorithm}
    \caption{Streaming algorithms for incoming walks.
    }
    \Input{Temporal graph $\tg=(V,\tge,\delta)$, function $\Phi_{in}$}
    \Output{Matrix $\WWalkIn$}
    \ForAll(\note*[f]{in chronological order}){$(u,v,t)\in \tge$}{\label{alg:streaming_algorithm:start_part1}
        \If{$\WWalkIn(v,t+\delta)$ not initialised}{$\WWalkIn(v,t+\delta)=0$}
        $\WWalkIn(v,t+\delta)\pluseq 1$\;
        \ForAll{$t'$ with $\WWalkIn(u,t')>0$}{
            \If{$t \geq t'$}{\label{alg:streaming_algorithm:timecheck}
                $\WWalkIn(v,t+\delta) \pluseq \WWalkIn(u,t') \cdot\Phi_{in}(t',t)$
            }
        } 
    }
\end{algorithm}
\begin{theorem}\label{theorem:streaming_runningtime}
     Let $\tg=(V,\tge,\delta)$ with $\delta>0$. 
     \Cref{alg:streaming_algorithm} computes $\WWalkIn$ correctly, and has a running time in $\mathcal{O}(|\tge|\cdot \tau^-_{max})$ and a space complexity in $\mathcal{O}(|V|\cdot \tau^-_{max})$,  with $\tau^-_{max}$ being the maximal size of the set of availability times of edges arriving at a node over all nodes.
\end{theorem}
Counting the matrix $\WWalkOut$ for the outgoing walks can be done symmetrically to \Cref{alg:streaming_algorithm} with equal time and space complexity, where we replace $\tau^-_{max}$ by $\tau^+_{max}$, i.e., the maximal number of distinct availability times of edges leaving a node over all nodes.
Hence, the total running time for both directions is in $\mathcal{O}(|\tge|\cdot \max\{\tau^-_{max}, \tau^+_{max}\})$ and the space complexity $\mathcal{O}(|V|\cdot \max\{\tau^-_{max}, \tau^+_{max}\})$.

\subsection{Computing the Temporal Walk Centrality from the Matrices}

After obtaining the matrices $\WWalkIn$ and $\WWalkOut$, we compute the walk centrality for all nodes.
At each node $v$ in $V$, for all pairs of arrival and starting times $t_{a}$ and $t_s$ with $t_a\leq t_s$ the function $\Phi_m(t_{a}, t_{s})$ must be evaluated. 
Hence, for each $v\in V$ we have to evaluate $\Phi_m$ at most $\tau^-(v)\cdot \tau^+(v)$ times. 
Under the assumption that we can evaluate $\Phi_m$ in constant time, the total running time is in $\mathcal{O}(|V|\cdot \tau^-_{max}\cdot \tau^+_{max})$.
Alternatively, we can observe that the running time is at most quadratic in the number of temporal edges.
In case that $\Phi_m(t_{a}, t_{s})=1$ for all $t_a,t_s\in\mathbb{N}$, we can compute the centrality of all nodes in time linear in the number of temporal edges.
\Cref{alg:matrixsum} iterates over all time points $T(v)$ in increasing order and iteratively sums up the number of incoming walks (line \ref{alg:matrixsum:sum}f.).
We multiply the total incoming weight with the current outgoing weight and add the result to the centrality value of $v$.
\begin{theorem}\label{theorem:centrality_runningtime}
    For $\Phi_m(t_1,t_2)=1$ and $t_1,t_2\in\mathbb{N}$, 
    \Cref{alg:matrixsum} computes the walk centrality from $\WWalkIn$ and $\WWalkOut$ in $\mathcal{O}(|\tge|)$ time.
\end{theorem}
\begin{algorithm}\label[algorithm]{alg:matrixsum}
    \caption{Computing temporal walk centrality for $\Phi_m(t_{a}, t_{s})=1$.}
    \label{alg:compute_walkcentrality}
    \Input{$\WWalkIn$ and $\WWalkOut$}
    \Output{Walk centrality $C(v)$ for all $v\in V$}
    \BlankLine
    \ForAll{$v\in V$}{
        $C(v)\gets 0$, $in_{sum} \gets 0$ \;
        \ForAll{$t\in T(v)$ in increasing order}{
            $in_{sum}\pluseq \WWalkIn(v,t)$\; \label{alg:matrixsum:sum}
            $C(v)\pluseq \WWalkOut(v,t)\cdot in_{sum}$
        }
    }
\end{algorithm}

\section{Experiments}\label{sec:experiments}
We discuss the following research questions:
\begin{itemize}[noitemsep,topsep=0pt]
    \item[\textbf{Q1.}] \textbf{Efficiency and Scalability:} How do our algorithms for computing the temporal walk centrality differ in terms of running time in practice? Do they scale to large networks?
    \item[\textbf{Q2.}] \textbf{Accuracy of \textsc{Approx}:} How is the accuracy of \textsc{Approx} compared to the exact results?
    \item[\textbf{Q3.}] \textbf{Effect of the Parameters:} How do the choices of the parameters affect the temporal walk centrality? 
    \item[\textbf{Q4.}] \textbf{Node Rankings:}  How do the rankings by temporal walk centrality compare to other temporal centrality measures?
\end{itemize}

\noindent
\textbf{Data Sets: }
We use fourteen real-world temporal graph data sets:  
(1)~\emph{Hospital} contains the contacts between hospital patients and medical personal~\cite{vanhems2013estimating}. 
(2)~\emph{HTMLConf} is a contact network of visitors of a conference~\cite{Isella2011}. %
(3)~\emph{Highschool} is a contact network of students over seven days~\cite{mastrandrea2015contact}. 
(4) \emph{College} is based on an online social network used by students~\cite{opsahl2009clustering,panzarasa2009patterns}.
(5) \emph{Infectious} represents face-to-face contacts of visitors of an exhibition~\cite{Isella2011}.
(6) \emph{Facebook} is a subset of the activity of a Facebook community~\cite{viswanath2009evolution}. 
(7) \emph{Enron} is an email network between employees of a company~\cite{klimt2004enron}. %
(8) \emph{AskUbuntu} is a network of interactions on the stack exchange website \emph{Ask Ubuntu}~\cite{paranjape2017motifs}.
(9) \emph{Digg} is a social network in which nodes represent persons and edges friendships. The time stamps indicate when friendships were formed~\cite{hogg2012social}. 
(10) \emph{Epinion} is based on a network of the product rating website \emph{Epinions}~\cite{richardson2003trust}.
(11) \mbox{\emph{WikiTalk}} is a social network based on the user pages of the \emph{Wikipedia} website, where nodes represent users and edges messages on the user page~\cite{SunKS16}.
(12) \emph{Wikipedia} is based on \emph{Wikipedia} pages and hyperlinks between them~\cite{mislove2009online}.
(13) \emph{Youtube} is a social network on a video platform~\cite{MisloveMGDB07}.
(14) \emph{Delicious} is based on a network of a bookmark website~\cite{wetzker2008analyzing}. 
\Cref{table:datasets_stats2} shows the properties and statistics of the data sets.
The transition time $\delta$ is one for all data sets.
\begin{table}
    \centering
    \caption{Statistics of the data sets with $G=\mathit{DL}(\tg)$.
        The type is either undirected ($u$) or directed ($d$).} 
    \label{table:datasets_stats2}
    \resizebox{1\linewidth}{!}{ \renewcommand{\arraystretch}{0.85} \setlength{\tabcolsep}{2.7pt}
        \begin{tabular}{lcrrrrrrrr}\toprule
            \multirow{4}{0.5cm}{\vspace*{4pt}\textbf{Data~set}\vspace*{4pt}}&\multicolumn{8}{c}{\textbf{Properties}}\\
            \cmidrule{2-9}
            \textbf{ }         &  Type    &   $|V(\tg)|$         & $|E(\tg)|$  &  $|{T}(\tg)|$ & $\tau^-_{max}$& $\tau^+_{max}$  &
            $|V(G)|$ & $|E(G)|$\\ \midrule
            \emph{Hospital}    & $u$ & $75$          & $32\,424$       & $9\,453$      & $2\,902$ & $2\,902$    & $64\,848$      & $62\,314\,564$  \\ 
            \emph{HTMLConf}    & $u$ & $113$         & $20\,818$       & $5\,246$      & $1\,390$ & $1\,390$    & $41\,636$      & $13\,535\,789$  \\
            \emph{Highschool}  & $u$ & $1\,894$      & $188\,508$      & $7\,375$      & $3\,500$ & $3\,500$    & $377\,016$     & $342\,993\,330$ \\  
            \emph{College}     & $d$ & $1\,899$      & $59\,835$       & $58\,911$     & $1\,539$ & $1\,539$    & $59\,835$      & $4\,039\,885$  \\
            \emph{Infectious}  & $u$ & $10\,972$     & $415\,912$      & $76\,943$     & $393$    & $393$       & $831\,824$     & $56\,569\,497$  \\ 
            \emph{Facebook}    & $d$ & $63\,731$     & $817\,035$      & $333\,924$    & $455$    & $455$       & $817\,035$     & $8\,051\,691$  \\ 
            \emph{Enron}       & $d$ & $87\,101$     & $1\,134\,046$   & $213\,167$    & $6\,033$ & $5\,353$    & $1\,134\,046$  & $361\,625\,773$  \\ 
            \emph{AskUbuntu}   & $d$ & $159\,316$    & $964\,437$      & $257\,079$    & $837$    & $2\,358$    & $257\,305$     & $2\,967\,386$  \\  
            \emph{Digg}        & $d$ & $279\,630$    & $1\,731\,652$   & $6\,865$      & $1\,328$ & $330$       & $1\,731\,652$  & $94\,858\,234$  \\ 
            \emph{Epinion}     & $d$ & $755\,760$    & $13\,668\,320$  & $501$         & $181$    & $498$       & $13\,668\,281$ & $94\,633\,962$  \\
            \emph{WikiTalk}    & $d$ & $1\,420\,367$ & $4\,641\,928$   & $3\,442\,682$ & $16\,213$&$1\,092\,785$& $4\,641\,928$  & $1\,696\,871\,574$ \\ 
            \emph{Wikipedia}   & $d$ & $1\,870\,709$ & $39\,953\,145$  & $2\,198$      & $1\,931$ & $489$       & $39\,953\,145$ & $2\,763\,845\,227$\\  
            \emph{Youtube}     & $d$ & $3\,223\,585$ & $9\,375\,374$   & $203$         & $191$    & $203$       & $9\,375\,374$  & $4\,410\,951\,091$  \\  
            \emph{Delicious}   & $d$ & $4\,512\,099$ & $219\,532\,884$ & $1\,583$      & $1\,583$ & $1\,317$    & $219\,532\,884$ & $83\,533\,929\,266$  \\  
            \bottomrule
        \end{tabular}
    }
\end{table}
\begin{table}
    \centering
    \caption{Running times in seconds (Oot--out of time, Oom--out of memory).}
    \label{table:runningtimes}
    \resizebox{0.8\linewidth}{!}{ \renewcommand{\arraystretch}{0.9}
        \begin{tabular}{lccrrrr}\toprule
            \multirow{4}{0.5cm}{\vspace*{4pt}\textbf{Data~set}\vspace*{4pt}}
            & \multicolumn{2}{c}{\textsc{Stream} with various $\Phi$} & \textsc{DlgMa} & \multicolumn{3}{c}{\textsc{Approx} for various $\varepsilon$}\\
            \cmidrule{2-3}\cmidrule{5-7}
            \textbf{ }               & $\Phi_\alpha$ & $\Phi_t$  &    & $0.1$ & $0.001$ & $0.00001$ \\\midrule
            \emph{Hospital}   & 2.50     & 3.04      & Oot & 18.32  & 18.69  & 19.00 \\ 
            \emph{HTMLConf}   & 0.56     & 0.58      & 5\,213.17 & 3.35   & 3.46   & 3.53 \\ 
            \emph{Highschool} & 20.98    & 23.24     & Oot & 117.95 & 123.67 & 128.44 \\ 
            \emph{College}    & 0.17     & 0.21      & Oot & 1.05   & 1.06   & 1.07 \\ 
            \emph{Infectious} & 1.29     & 1.33      & Oot & 12.75  & 13.06  & 13.22 \\ 
            \emph{Facebook}   & 0.57     & 0.64      & Oot & 4.23   & 4.30   & 4.30 \\  
            \emph{Enron}      & 11.31    & 15.09     & Oot & 124.13 & 129.33 & 131.94 \\ 
            \emph{AskUbuntu}  & 0.21     & 0.28      & Oot & 1.06   & 1.07   & 1.08 \\ 
            \emph{Digg}       & 1.70     & 1.81      & Oot & 41.95  & 42.48  & 42.90 \\ 
            \emph{Epinion}    & 2.62     & 2.71      & Oot & 41.31  & 41.38  & 41.83 \\ 
            \emph{WikiTalk}   & 126.42   & 226.25    & Oom & Oom & Oom & Oom \\ 
            \emph{Wikipedia}  & 190.11   & 200.81    & Oom & Oom & Oom & Oom \\ 
            \emph{Youtube}    & 9.63     & 9.91      & Oom & Oom & Oom & Oom \\ 
            \emph{Delicious}  & 1\,851.68& 1\,928.51 & Oom & Oom & Oom & Oom \\
            \bottomrule
        \end{tabular}
    }
\end{table}
\begin{table}
    \centering
    \caption{Mean relative errors of \textsc{Approx} for various $\varepsilon$.}
    \label{table:accuracy}
    \resizebox{0.65\linewidth}{!}{ \renewcommand{\arraystretch}{0.9}\setlength{\tabcolsep}{10.8pt}
        \begin{tabular}{lccc}\toprule
            \multirow{4}{0.5cm}{\vspace*{4pt}\textbf{Data~set}\vspace*{4pt}}
            & \multicolumn{3}{c}{\textsc{Approx} with various $\varepsilon$}\\
            \cmidrule{2-4}
            \textbf{ }        & $0.1$    & $0.001$  & $0.00001$ \\\midrule
            \emph{Hospital}   & 2.20E-08 & 1.38E-10 & 3.89E-12 \\ 
            \emph{HTMLConf}   & 6.04E-08 & 1.08E-09 & 1.69E-12 \\ 
            \emph{Highschool} & 1.64E-09 & 9.13E-12 & 4.63E-14 \\ 
            \emph{College}    & 4.00E-08 & 8.78E-11 & 3.87E-12 \\ 
            \emph{Infectious} & 1.11E-09 & 2.71E-12 & 1.26E-13 \\ 
            \emph{Facebook}   & 3.29E-12 & 9.08E-15 & 9.08E-15 \\ 
            \emph{Enron}      & 3.44E-09 & 7.13E-12 & 9.35E-14 \\ 
            \emph{AskUbuntu}  & 5.42E-10 & 5.71E-12 & 5.50E-14 \\ 
            \emph{Digg}       & 5.20E-12 & 7.30E-14 & 1.10E-15 \\
            \emph{Epinion}    & 2.89E-16 & 2.89E-16 & 1.22E-16 \\ 
            \bottomrule
        \end{tabular}
    }
\Description{The accuracies of the approximation algorithm.}
\end{table}
\begin{figure*}[ht]
    \centering
    \begin{subfigure}{.25\linewidth}
        \centering
        \includegraphics[width=1\linewidth]{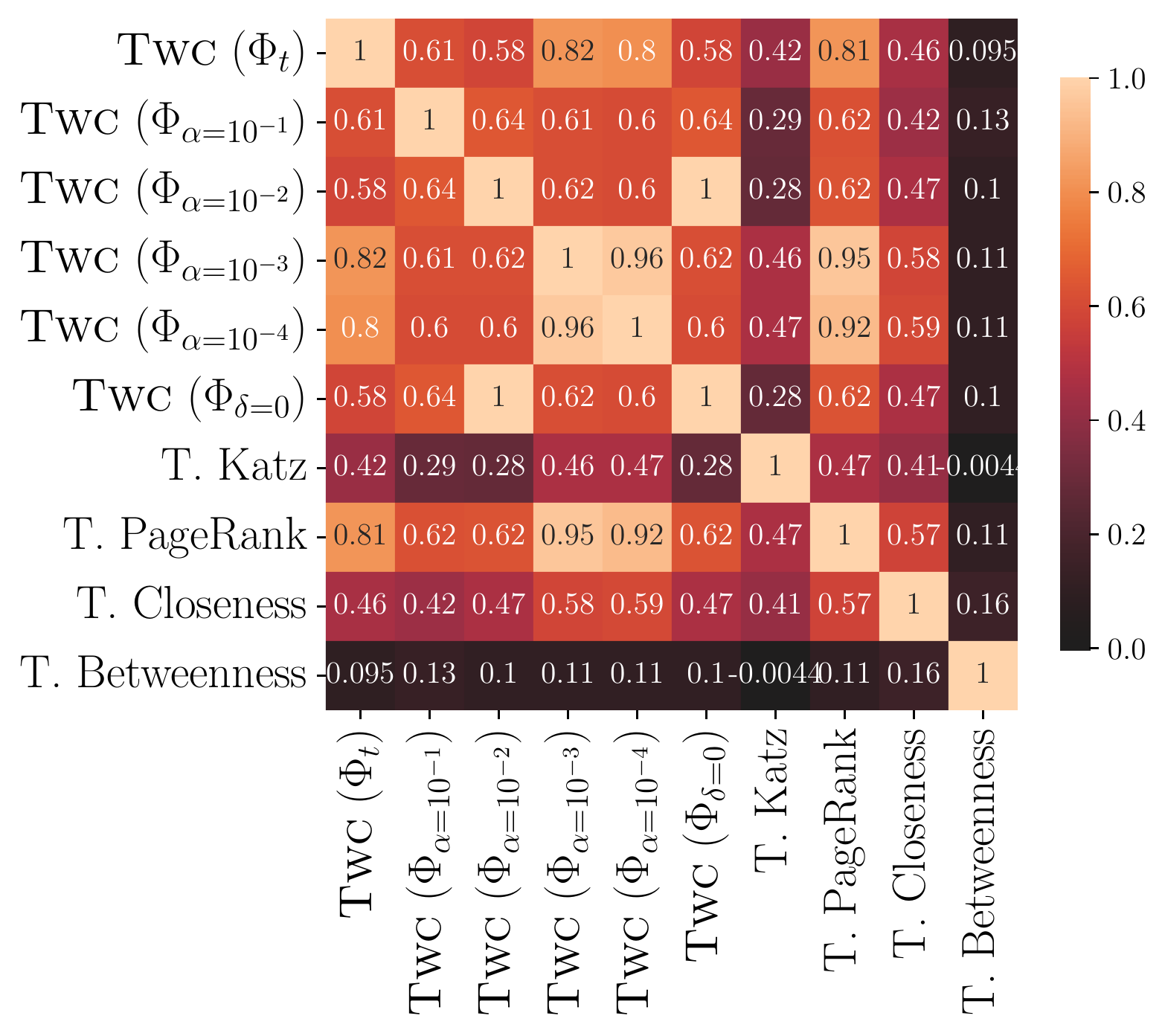}
        \caption{\emph{HTMLConf}}
    \end{subfigure}\hfil%
    \begin{subfigure}{.25\linewidth}
        \centering
        \includegraphics[width=1\linewidth]{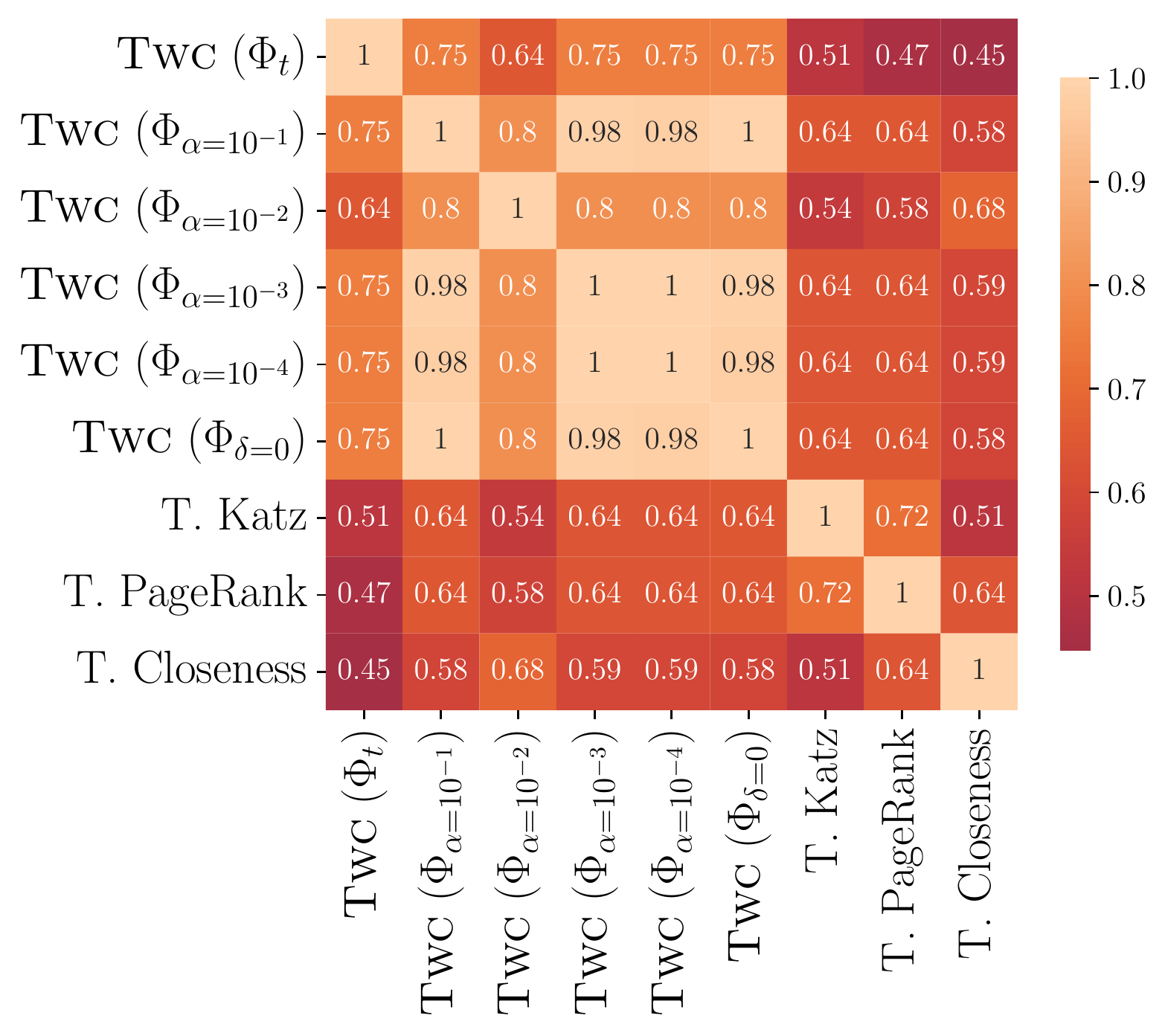}
        \caption{\emph{Facebook}}
    \end{subfigure}\hfil%
    \begin{subfigure}{.25\linewidth}
        \centering
        \includegraphics[width=1\linewidth]{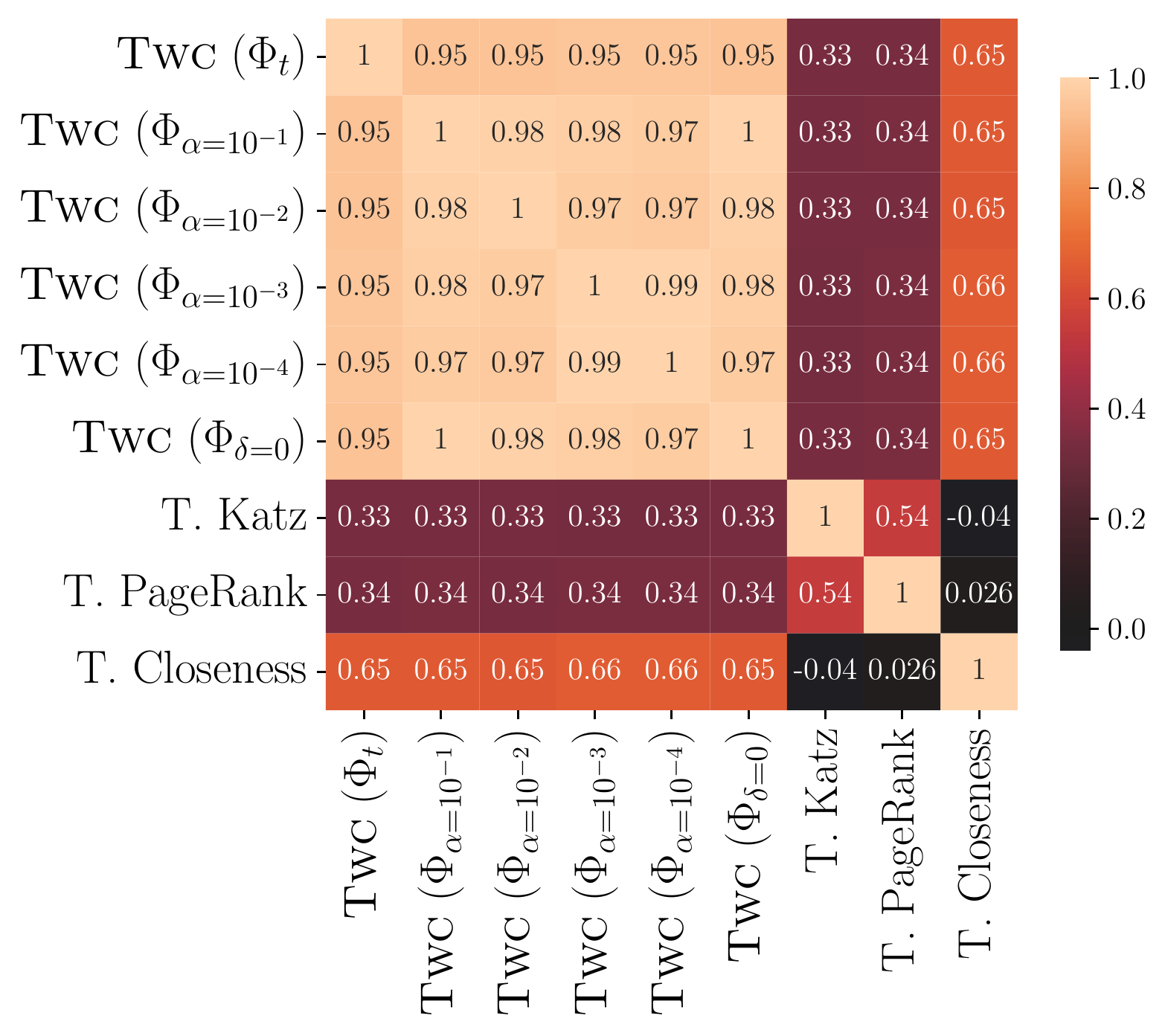}
        \caption{\emph{Enron}}
    \end{subfigure}\hfil%
    \begin{subfigure}{.25\linewidth}
        \centering
        \includegraphics[width=1\linewidth]{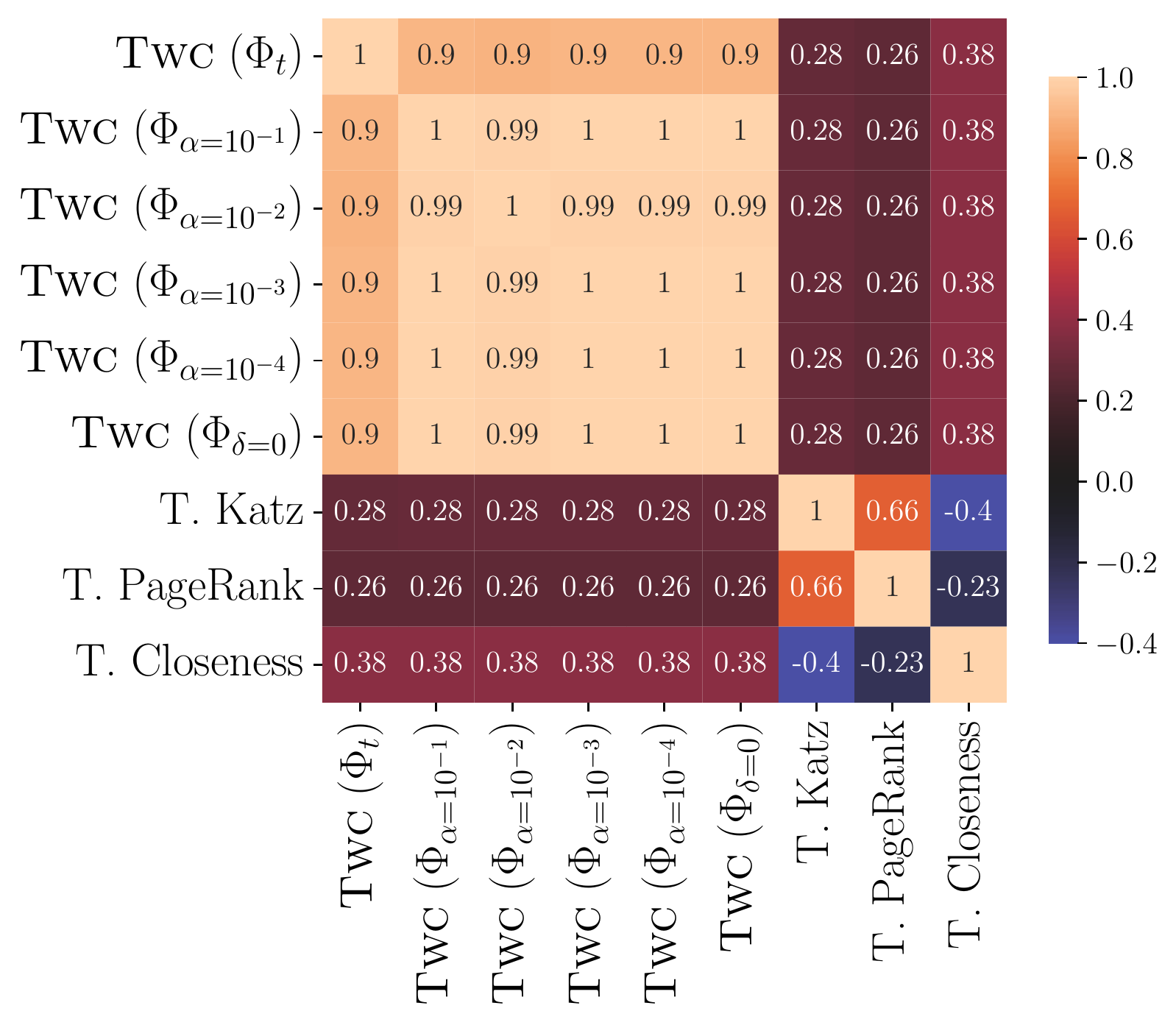}
        \caption{\emph{AskUbuntu}}
    \end{subfigure}\hfil%
    \vspace{-2mm}
    \caption{Kendall rank correlation between the rankings computed using different variants of the temporal walk centrality and other temporal centrality measures.}
    \label{fig:correlation2}
    \Description{Correlation matrices showing the Kendal rank correlation.}
\end{figure*}
\subsection{Algorithms and Experimental Protocol}
We implemented the following algorithms in C++ using the GNU CC Compiler 10.3.0 and the \texttt{Eigen} library for matrix operations.
\begin{itemize}[noitemsep,topsep=0pt] 
    \item \textsc{Stream} is the implementation of \Cref{alg:streaming_algorithm}. 
    \item \textsc{DlgMa} uses the directed line graph expansion (DLG) and matrix inversion (\Cref{sec:dlg}).
    \item \textsc{Approx} is the DLG-based approximation (\Cref{alg:dlg:approx}).
\end{itemize}
All experiments ran on a computer cluster. Each experiment had an exclusive node with an Intel(R) Xeon(R) Gold 6130 CPU @ 2.10GHz and 192 GB of RAM. The time limit was set to two hours.
The source code is available at \url{https://gitlab.com/tgpublic/twc}.

\subsection{Results and Discussion}
\noindent\textbf{Q1. Efficiency and Scalability:} 
First, we evaluated the running times of our algorithms.
For \textsc{Stream}, we use both $\Phi_\alpha(t_1,t_2)=\alpha$ and $\Phi_t(t_1,t_2)=\frac{1}{1+t_2-t_1}$ to compute two variants of the temporal walk centrality. For the other algorithms, we use $\Phi_\alpha$. In all cases, we set $\alpha=0.001$.
We set $\Phi_m(t_1,t_2)=1$ in case of $\Phi_\alpha$, and $\Phi_m(t_1,t_2)=\Phi_t(t_1,t_2)$ otherwise.
\Cref{table:runningtimes} shows the results.
For \textsc{Stream} the running time is lowest for all data sets.
It is at least five times faster than the approximation \textsc{Approx},
and several orders of magnitude faster than \textsc{DlgMa}.
In the case of $\Phi_t$, compared to $\Phi_\alpha$, the running times increase. 
The increase is less than $10\%$ for half of the data sets, and on average only $17.3\%$.
For \emph{WikiTalk} the increase is the most with $78.9\%$. The reason are the large values of the maximal number of starting or arrival times $\tau^-_{max}$ and $\tau^+_{max}$ (see \Cref{table:datasets_stats2}). 
\emph{Youtube} has small values for $\tau^-_{max}$ and $\tau^+_{max}$, and hence, the computations of \textsc{Stream} for both variants of $\Phi$ are much faster compared to \emph{WikiTalk}, even though \emph{Youtube} has more nodes and edges.
\textsc{DlgMa} could only compute the result for the \emph{HTMLConf} data set in the given time limit of two hours. The reason is that the input matrices are large and the computed inverse matrices are not always sparse. However, the approximation algorithm \textsc{Approx} computed the centrality values efficiently.
The very large DLGs for \emph{WikiTalk}, \emph{Wikipedia}, \emph{Youtube}, and \emph{Delicious} (see \Cref{table:datasets_stats2}) lead to 
out-of-memory errors during the computation of \textsc{DlgMa}.
Even though the DLG is often much smaller than the theoretical maximal size, the sizes of the DLGs can be a bottle-neck of the DLG-based algorithms. 
For example, for the \emph{Enron} data set \textsc{Approx} uses $59.7$ GB memory, were \textsc{Stream} only needs $0.58$ GB (see also \Cref{table:memory} in the appendix).
In the case that $\delta>0$ and we only have to consider strict temporal walks, \textsc{Stream} shows very good scalability.
Even for the large data sets \emph{Wikipedia} and \emph{Delicious} with around $40$ and $220$ million edges, respectively, the computations are time- and space-efficient.
\Cref{table:memory} in \Cref{appendix:memory} shows the memory usage.
\noindent\textbf{Q2. Accuracy of \textsc{Approx}:}
We evaluated the accuracy of our approximation algorithm \textsc{Approx} for $\alpha=0.001$.
\Cref{table:accuracy} shows the mean relative errors for $\varepsilon\in\{0.1,0.001,0.00001\}$ compared to the exact results computed with \textsc{Stream} for all data sets for which the DLG could be computed. Let $W=\{v\in V\mid C(v)\neq 0\}$ and $\hat{C}(v)$ be the approximated value for $v\in V$, we report $\frac{1}{|W|}\sum_{v\in W}\frac{|C(v)-\hat{C}(v)|}{C(v)}$. %
For all $\varepsilon$, the errors are insignificant and very low. The error decreases for smaller $\varepsilon$ as expected.
A smaller value of $\alpha$ leads to fast convergence.
In conclusion, these results show that \textsc{Approx} is highly accurate while being efficient (see \textbf{Q1}). 
\noindent\textbf{Q3. Effect of the Parameters:}
We computed the node rankings using temporal walk centrality for $\Phi_\alpha$, $\alpha \in \{0.1,0.01,0.001,0.0001\}$, and for $\Phi_t$. 
Furthermore, we set the transition time $\delta=0$ and computed the temporal walk centrality with $\Phi_\alpha$ for $\alpha=0.001$.
We measured the pairwise Kendall rank correlations (Kendall's $\tau$ coefficient) between the rankings \cite{kendall1938new}.
The Kendall rank correlation coefficient is commonly used for determining the relationship between centrality measures \cite{grando2016analysis}. The correlation coefficient takes on values between one and minus one, where values close to one indicate similar rankings, close to zero no correlation, and close to minus one a strong negative correlation.
\Cref{fig:correlation2} shows correlation matrices for \emph{HTMLConf}, \emph{Facebook}, \emph{Enron}, and \emph{AskUbuntu}. 
We observed similar results for the other data sets.
\textsc{Twc} denotes the different rankings computed with the variations of the temporal walk centrality. 
The correlation between the rankings using temporal walk centrality with $\Phi_\alpha$ is often strong for different $\alpha$.
The influence of $\alpha$ seems to be limited for larger data sets when we consider the complete rankings of all nodes, because the majority of nodes obtain similar positions. In case that we only consider a fraction of the nodes with the highest centralities, the impact of $\alpha$ is stronger (see \Cref{fig:correlation2_top1p} in \Cref{appendix:additional_results}) because the ratio of differently ranked nodes increases. 
Using a zero transition time of $\delta=0$ did not lead to different rankings compared to $\delta=1$.
The correlation between the rankings using $\Phi_t$ and $\Phi_\alpha$ is  often weaker than any of the correlations between the other \textsc{Twc} rankings.
Hence, using the waiting time-based weighting can lead to temporal walk centrality rankings different from using distance-based weighting.

\noindent\textbf{Q4. Node Rankings:}
We used the temporal Page\-Rank ($\alpha=0.99$, $\beta = 0.5$), temporal closeness, temporal Katz centrality (constant weighting with $\beta=0.01$), and temporal betweeness to compute the node rankings for the \emph{HTMLConf}, \emph{Facebook}, \emph{Enron}, and \emph{AskUbuntu} data sets. Due to the high running times of the temporal betweenness, the results are only computed for the \emph{HTMLConf} data set.
We report the results in the correlation matrices shown in \Cref{fig:correlation2}.
The correlations between the variants of the temporal walk centrality and the other temporal centrality measures are between -0.004 and 0.68, with the lowest values for temporal betweeness.
Hence, the rankings of the other temporal centrality measures have only a weak association with the rankings obtained using the temporal walk centrality.
This is expected, as the other centrality measures are not designed for ranking vertices according to their importance in information spreading.

\section{Conclusion and Future Work} 
We introduced the temporal walk centrality for temporal networks, which captures the intuition of important nodes capable of obtaining and distributing information efficiently. We illustrated how the temporal walk centrality can identify nodes that are crucial for the dissemination of information. 
We theoretically and experimentally showed that temporal walk centrality can be computed efficiently and with high accuracy in the case of our approximation. Moreover, our streaming algorithm scales to very large temporal networks. 
In future work, more general weight functions for temporal walks could be studied, which not only depend on two points in time but, e.g., also on the number and availability times of incident edges of a node in the walk. 
Suitable weight functions can allow a probabilistic interpretation of the temporal walk centrality.
\begin{acks}
    This work is funded by the Deutsche Forsch\-ungsgemeinschaft (DFG, German Research Foundation) under Germany's Excellence Strategy -- EXC-2047/1 -- 390685813.
    Nils Kriege has been supported by the Vienna Science and Technology Fund (WWTF) through project VRG19-009.
\end{acks}

\appendix
\section{Notation}\label{appendix:notation}
\Cref{table:notation} shows our commonly used symbols and their definitions.
\begin{table}[ht]
    \caption{Commonly used notations}
    \label{table:notation}
    \centering%
    \resizebox{1\linewidth}{!}{\renewcommand{\arraystretch}{1.0}%
        \begin{tabular}{c@{\hspace{3mm}}l@{}}%
            \toprule
            \textbf{Symbol} & \textbf{Definition}
            \\\midrule
            $\tg=(V,\tge,\delta)$                   & Temporal graph $\tg$ with nodes $V$ and temp. edges $\tge$ \\
            $\delta$                                & Global transition time\\
            $e=(u,v,t)$                             & Temporal $(u,v)$-edge at time $t$\\
            $T(v)$                                  & Set of availability times of edges incident to node $v$\\
            $T(\tg)$                                & Set of availability and arrival times in $\tg$\\
            $G=(V,E,w)$                             & Directed graph with edge weights $w\colon E \to \mathbb{R}$\\
            $\Phi:\mathbb{N}\times \mathbb{N}\rightarrow \mathbb{R}$  & Time depended weighting function\\
            $\WalkIn(v,t)$, $\WalkOut(v,t)$         & Temporal walks ending, starting at node $v$ and time $t$ \\
            $\WWalkIn(v,t)$, $\WWalkOut(v,t)$       & Total weight of temp.~walks in $\WalkIn(v,t)$, $\WalkOut(v,t)$\\
            $\tau^-(v)$, $\tau^+(v)$                & \# of distinct arrival, starting times at $v\in V$\\
            $\tau^-_{max},\tau^+_{max}$             & Max. \# of distinct arrival, starting times over $v\in V$\\
            $\varepsilon$                           & Error-tolerance parameter for approximation \\
            \bottomrule
    \end{tabular}}
\end{table}
\section{Omitted Proofs}\label{sec:proofs}
\begin{proof}[Proof of \Cref{lemma:dlg_walks}]
    Let $\omega$ be a walk in $\mathit{DL}(\tg)$. It directly follows from \Cref{def:dlg}, that $\Gamma(\omega)$ is a walk in $\tg$ that satisfies the time constraints. Vice versa, every temporal walk $\omega$ in $\tg$ corresponds to a sequence of edges, whose nodes are adjacent in $\mathit{DL}(\tg)$.
\end{proof}
\begin{proof}[Proof of \Cref{lemma:dlg_weights}]
    Let $\omega=\left(n^{t_1}_{v_1v_2}, n^{t_2}_{v_2v_3}, \ldots, n^{t_\ell}_{v_\ell v_{\ell+1}}\right)$. Application of edge weights yields $w_\Phi(\omega) = \prod_{i=1}^{\ell-1}\Phi(t_i+\delta,t_{i+1}) = \tau_\Phi(\Gamma(\omega))$.
\end{proof}
\begin{proof}[Proof of \Cref{theorem:streaming_runningtime}]
    Let $\omega_\ell$ be a walk of length $1\leq \ell\leq k$ arriving at vertex $v$.
    For $\ell=1$, the walk consists of a single edge $e_1=(u,v,t)$. \Cref{alg:streaming_algorithm} iterates over all edges, and therefore, also over $e_1$.
    Therefore, $\WWalkIn(v,t+\delta)$ will be initialized with one, and the walk is counted. 
    For $\ell>1$, we now have to check that we count only temporal walks. However, because the algorithm processes the edges in chronological order and due to line \ref{alg:streaming_algorithm:timecheck}, the edge can only extend walks that arrive not later at $u$ than $t$.
    Consider a walk $\omega_2=((u,w,t),(w,v,t'))$ of length $\ell=2$ arriving at $v$. The walk $((u,w,t))$ arriving at vertex $w$ was counted before because $t+\delta \leq t'$, hence $\WWalkIn(w,t+\delta)>0$. 
    Next, the walks from $w$ are added to the walks at $v$, weighted by $\Phi_{in}(t,t')$. 
    Consider the walk $\omega_{\ell}=(e_1,\ldots,e_{\ell}=(v_{\ell},v_{\ell+1},t_{\ell}))$.  
    By our assumption, it follows that the path $\omega_{\ell-1}=(e_1,\ldots,e_{\ell-1}=(v_{\ell-1},v_{\ell},t_{\ell-1}))$ of length $\ell-1$ arrived at time $t_{\ell-1}+\delta$ at $v_{\ell}$ and was counted correctly. 
    The algorithm adds the walks of length ${\ell-1}$ arriving at $v_{\ell}$ to the number of walks of length $\ell$ at vertex $v_{\ell+1}$ weighted by $\Phi_{in}(t_{\ell-1},t_{\ell})$.
    
    For the running time, \Cref{alg:streaming_algorithm} iterates over all $|\tge|$ edges in chronological order, and for each $e\in\tge$, it iterates over all entries of $\WWalkIn$ for which $\WWalkIn(e.u,t)>0$, i.e., at most $\tau^-_{max}$ rows. Therefore, the running time is in $\mathcal{O}(|\tge|\cdot \tau^-_{max})$, using a minimal perfect hash function for indexing the arrival times.
    For each vertex $v\in V$ and arrival time $t_a$, 
    we store the sum of the weighted walks arriving at $v$ at time $t_a$. 
\end{proof}
\begin{proof}[Proof of \Cref{theorem:centrality_runningtime}]
    \Cref{alg:matrixsum} iterates over all nodes and over all time stamps $t\in T(v)$.
    The sum of the time stamps over all nodes as well as the number of nodes itself are bounded by $|\tge|$.
\end{proof}

\section{Computation in Acyclic DLG}\label{sec:acyclic}
For the special case of acyclic graphs we can sum the weights of incoming and outgoing walks for all nodes in linear total time. The weight of walks starting at a node is the sum of the weighted walks starting at its outgoing neighbors.
\Cref{alg:dag_walk_count} implements weighted walk counting in a bottom-up fashion starting at the sinks and propagating weighted walk counts level-wise upwards.
\begin{lemma}
    The weighted temporal walks in a temporal graph $\tg=(V,\tge,\delta)$ with $\delta>0$ are counted exactly by \Cref{alg:dag_walk_count} in time and space $\mathcal{O}(|\tge|+e)$, where $e=|E(\mathit{DL}(\tg))|$.
\end{lemma}
We can count the incoming weighted walks for every node analogously, either by reversing all edges, or by considering the outgoing neighbors starting from the sources.

\begin{algorithm}
    \label[algorithm]{alg:dag_walk_count}
    \caption{Counting weighted outgoing walks in directed acyclic graphs.}
    \Input{Directed weighted acyclic graph $G=(V,E,w)$.}
    \Output{Weighted walk counts $\WWalkOutStatic(v)$ for all $v\in V$.}
    \BlankLine
    
    \lForAll(\note*[f]{Length $0$ walks}){$v \in V$} {
        $\WWalkOutStatic(v) \gets 1$
    }
    $S^0 \gets \{v \in V \mid d^+(v) = 0 \}$ \note*[r]{Find sinks}
    $i \gets 0$ \;
    \Repeat{$S^i \neq \emptyset$} {
        \ForAll{$v \in S^i$} {
            \ForAll{$u \in N^-(v)$} {
                $\WWalkOutStatic(u) \gets \WWalkOutStatic(u) + w(u,v)\cdot\WWalkOutStatic(v)$\; 
                $S^{i+1} \gets S^{i+1} \cup \{u\}$ \note*[r]{Avoid duplicates}
            }
        }
        $i \gets i + 1$ \;
    }
\end{algorithm}

\section{Additional Results}\label{appendix:additional_results}
This section provides additional experimental results. 

\subsection{Memory Usage}\label{appendix:memory}
\Cref{table:memory} shows the total memory usage of \textsc{Stream} and \textsc{Approx}.
For \textsc{Stream}, we report the results for $\Phi_\alpha$ with $\alpha=0.001$, and for \textsc{Approx} with $\epsilon=0.001$.
Note the much higher memory usage of \textsc{Approx} due to the computation of the DLG.

\subsection{Top-k Node Rankings}

\begin{figure*}[!t]
    \centering
    \begin{subfigure}{.4\linewidth}
        \centering
        \includegraphics[width=1\linewidth]{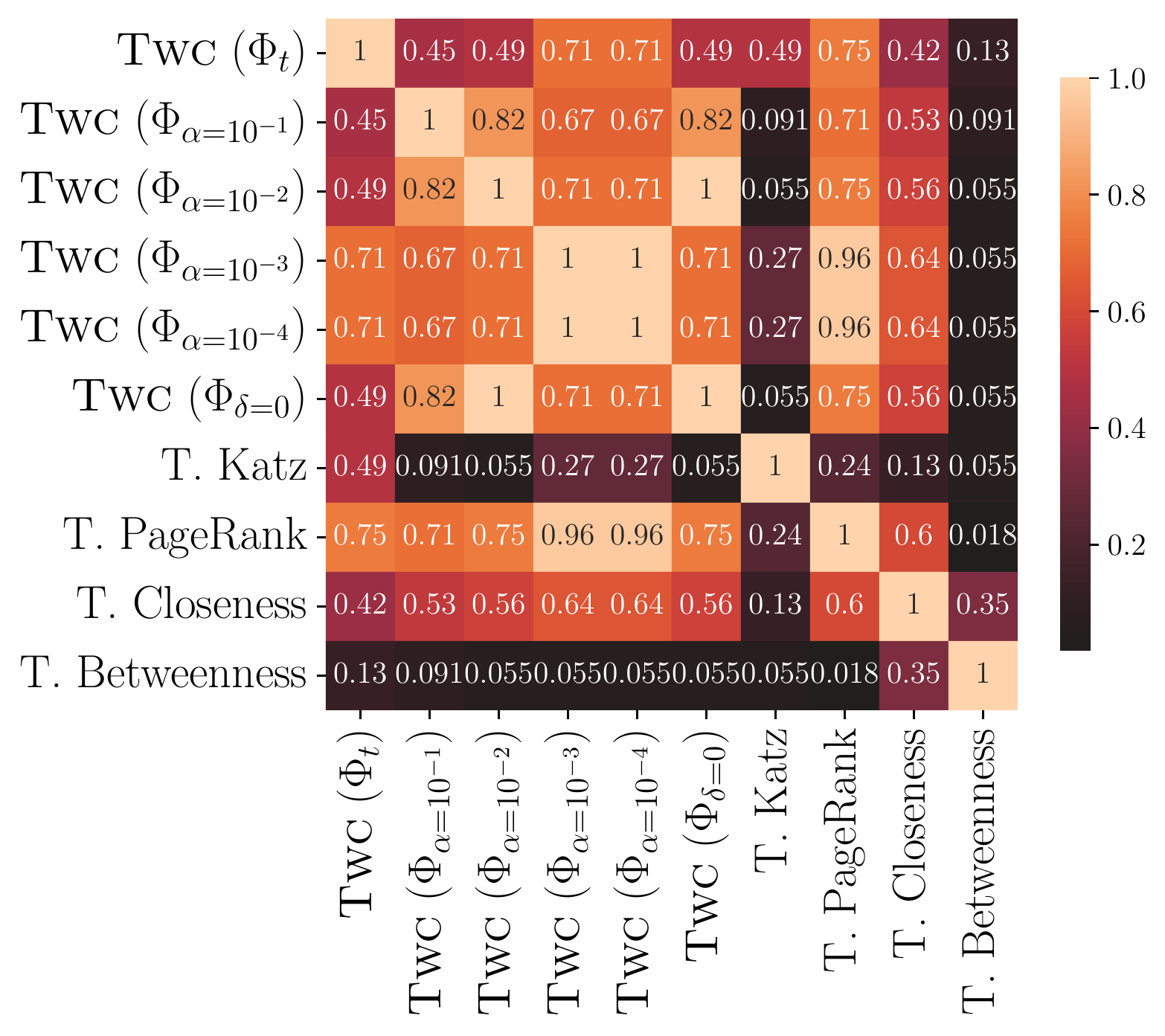}
        \caption{\emph{HTMLConf}}
    \end{subfigure}\hfil%
    \begin{subfigure}{.4\linewidth}
        \centering
        \includegraphics[width=1\linewidth]{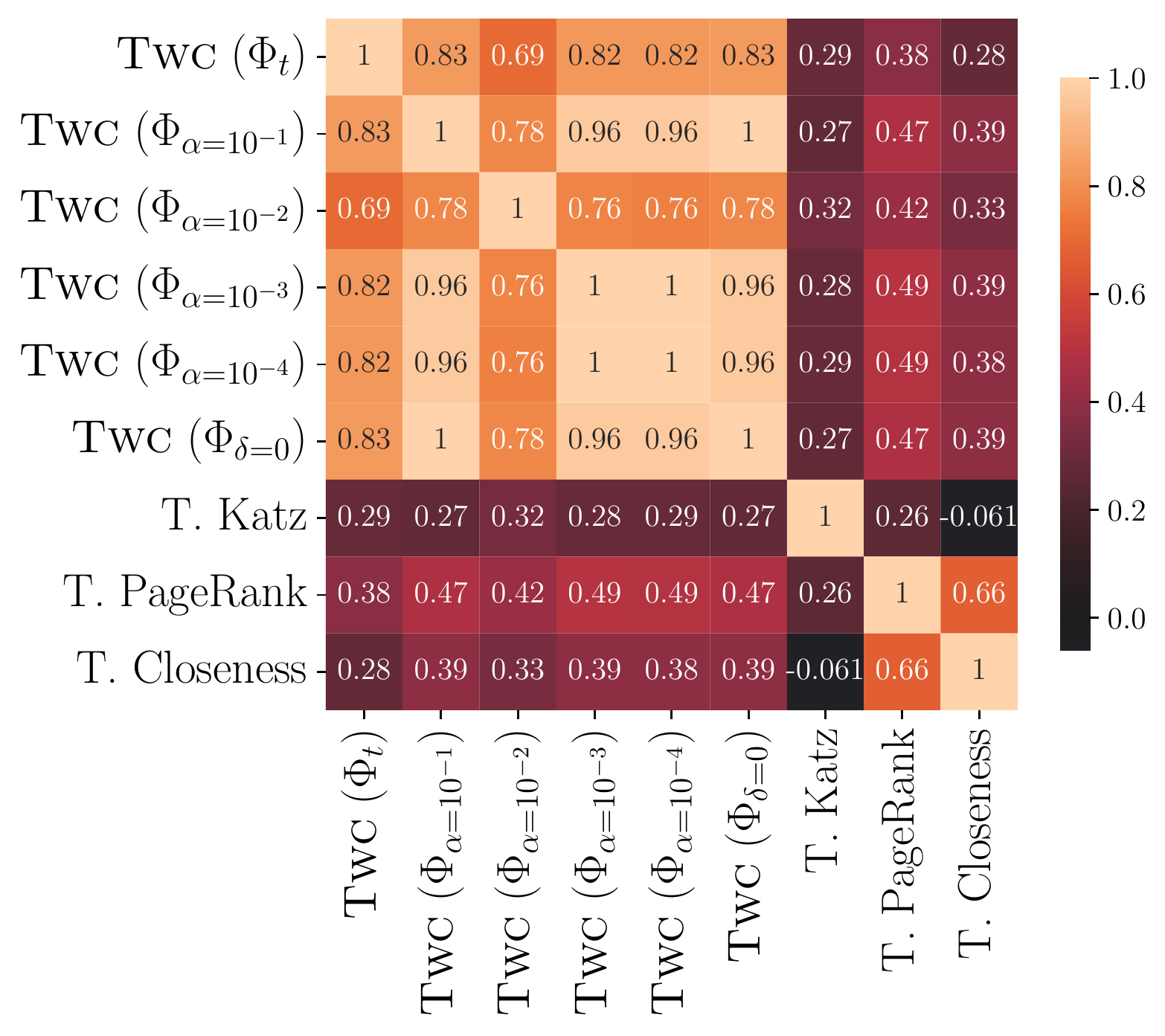}
        \caption{\emph{Facebook}}
    \end{subfigure}\\
    \begin{subfigure}{.4\linewidth}
        \centering
        \includegraphics[width=1\linewidth]{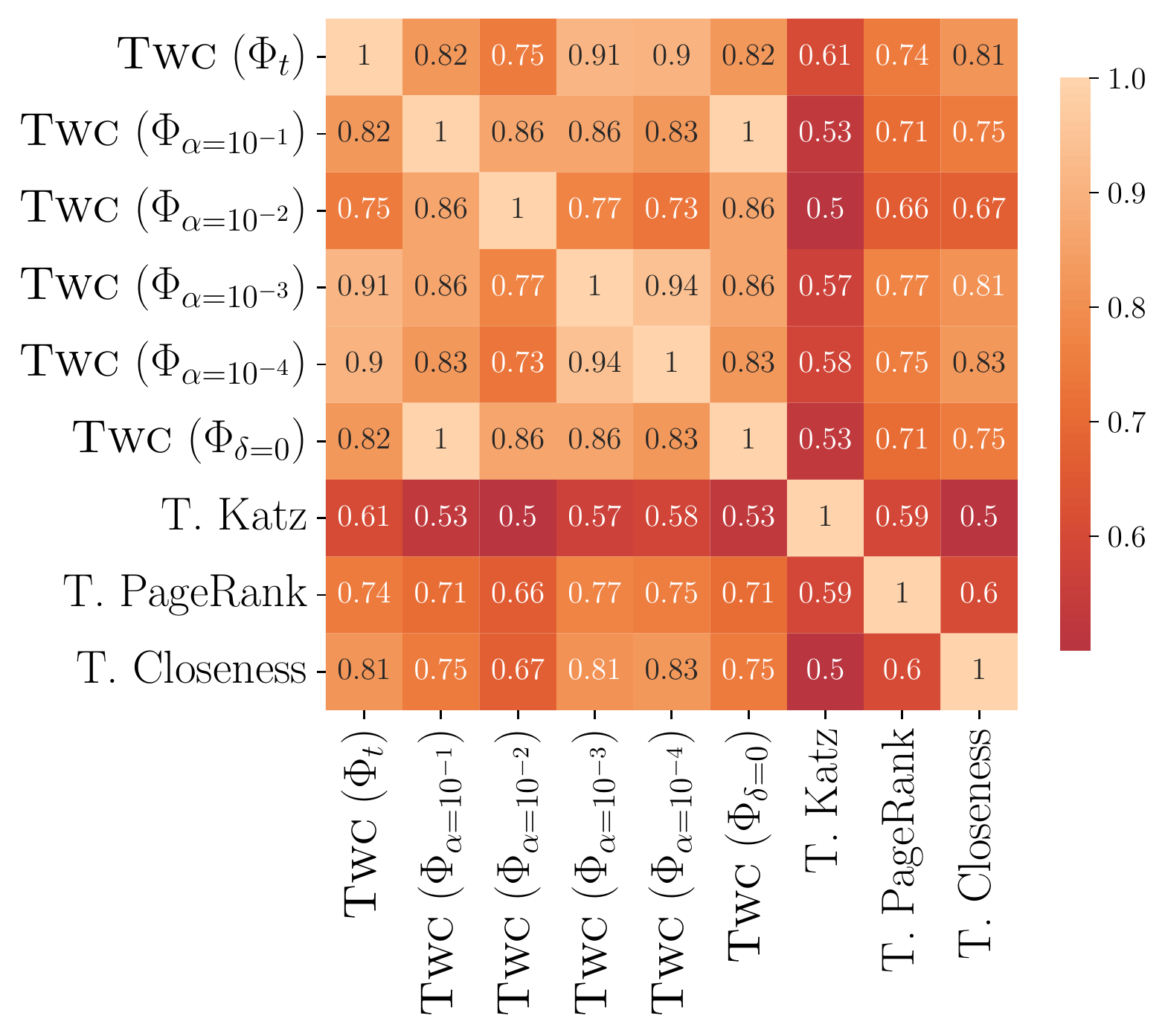}
        \caption{\emph{Enron}}
    \end{subfigure}\hfil%
    \begin{subfigure}{.4\linewidth}
        \centering
        \includegraphics[width=1\linewidth]{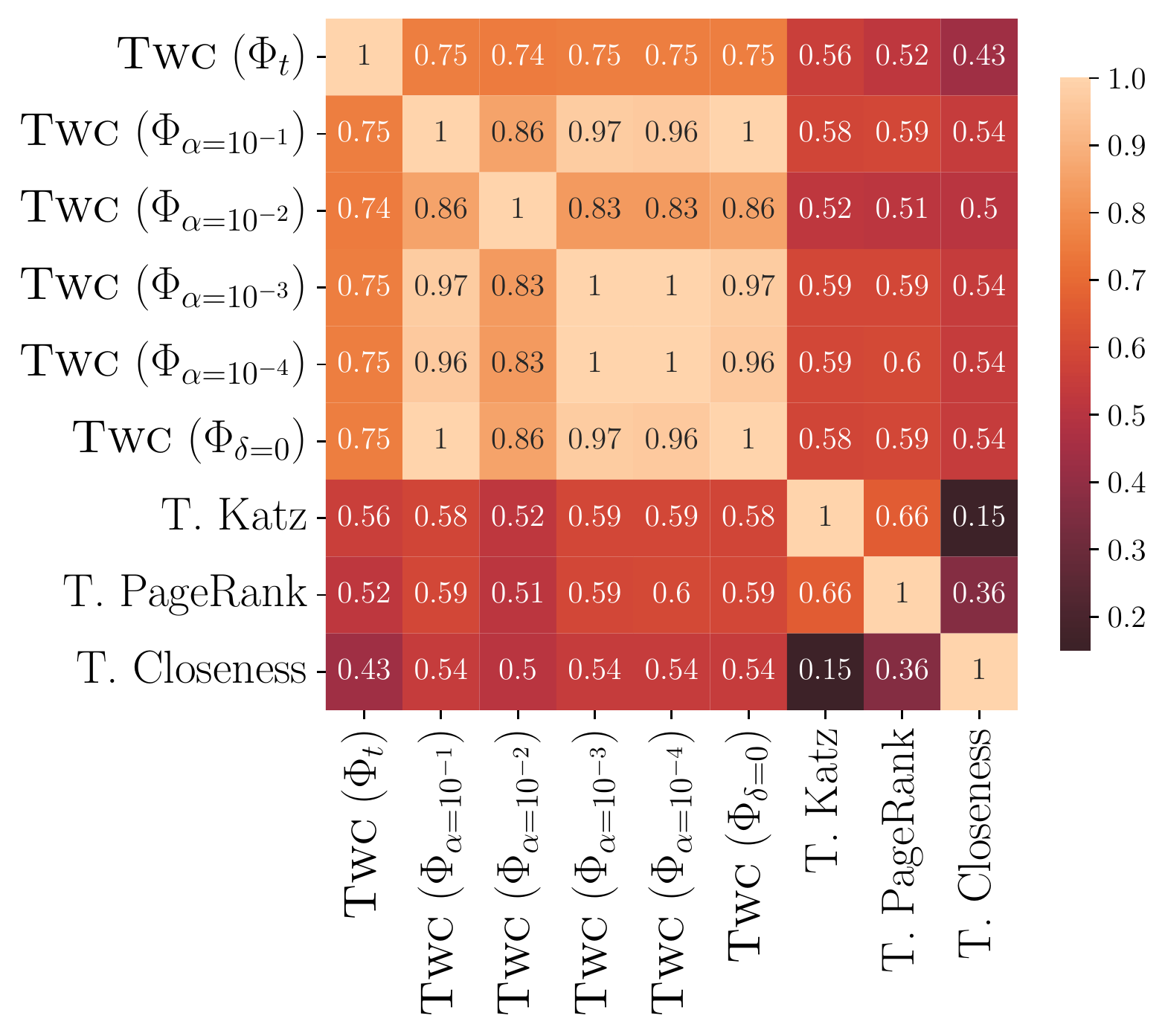}
        \caption{\emph{AskUbuntu}}
    \end{subfigure}
    \caption{Kendall rank correlation between the rankings computed using different variants of the temporal walk centrality and other temporal centrality measures using only the top-$k$, with $k=\lceil |V|\cdot p \rceil$ nodes with the highest centrality values with $p=0.1$ for \emph{HTMLConf} and $p=0.001$ for the other data sets.}
    \Description{Correlation matrices showing the Kendal rank correlation.}
    \label{fig:correlation2_top1p}
\end{figure*}

\Cref{fig:correlation2_top1p} shows the correlation matrices that are based on the top-$k$ rankings computed with the variants of \textsc{Twc}, temporal PageRank, temporal Katz, temporal betweenness, and temporal closeness centralities. The temporal betweenness is due to the long running times only computed for the \emph{HTMLConf} data set.
The value $k$ is chosen to be $10\%$ of the number of the vertices of the small \emph{HTMLConf} data set, and $0.1\%$ for the other, larger data set.
We chose a higher percentage for \emph{HTMLConf} due to the small number of vertices of the data set.
The impact of $\alpha$ in case of the top-$k$ rankings is higher, because compared to the complete rankings the ratio of nodes that change their position in the ranking is higher.

\begin{table}
    \centering
    \caption{Memory usage in GB (Oom--out of memory).}
    \label{table:memory}
    \resizebox{0.8\linewidth}{!}{ \renewcommand{\arraystretch}{0.9}\setlength{\tabcolsep}{20.8pt}
        \begin{tabular}{lrr}\toprule
            & \textsc{Stream} & \textsc{Approx} \\ \cmidrule{2-3}
            \textbf{Data set} & $\alpha=0.001$  & $\varepsilon=0.001$ \\\midrule
            \emph{Hospital}   &   0.034 & 10.152 \\ 
            \emph{HTMLConf}   &   0.023 &  2.295 \\ 
            \emph{Highschool} &   0.183 & 56.035 \\ 
            \emph{College}    &   0.033 &  0.737 \\ 
            \emph{Infectious} &   0.422 & 10.156 \\ 
            \emph{Facebook}   &   0.409 &  1.866 \\ 
            \emph{Enron-Rm}   &   0.582 & 59.676 \\ 
            \emph{AskUbuntu}  &   0.168 &  0.169 \\ 
            \emph{Digg}       &   0.883 & 17.129 \\ 
            \emph{Epinion}    &   6.718 & 23.331 \\ 
            \emph{WikiTalk}   &   2.221 & Oom \\ 
            \emph{Wikipedia}  &  20.330 & Oom \\ 
            \emph{Youtube}    &   5.126 & Oom \\ 
            \emph{Delicious}  & 106.999 & Oom \\ 
            \bottomrule
        \end{tabular}
    }
\end{table}


\begin{thebibliography}{58}
    
    %
    %
    %
    %
    %
    %
    %
    %
    %
    %
    %
    %
    %
    %
    %
    %
    
    \ifx \showCODEN    \undefined \def \showCODEN     #1{\unskip}     \fi
    \ifx \showDOI      \undefined \def \showDOI       #1{#1}\fi
    \ifx \showISBNx    \undefined \def \showISBNx     #1{\unskip}     \fi
    \ifx \showISBNxiii \undefined \def \showISBNxiii  #1{\unskip}     \fi
    \ifx \showISSN     \undefined \def \showISSN      #1{\unskip}     \fi
    \ifx \showLCCN     \undefined \def \showLCCN      #1{\unskip}     \fi
    \ifx \shownote     \undefined \def \shownote      #1{#1}          \fi
    \ifx \showarticletitle \undefined \def \showarticletitle #1{#1}   \fi
    \ifx \showURL      \undefined \def \showURL       {\relax}        \fi
    %
    %
    \providecommand\bibfield[2]{#2}
    \providecommand\bibinfo[2]{#2}
    \providecommand\natexlab[1]{#1}
    \providecommand\showeprint[2][]{arXiv:#2}
    
    \bibitem[\protect\citeauthoryear{Alman and Williams}{Alman and
        Williams}{2021}]%
    {AlmanW21}
    \bibfield{author}{\bibinfo{person}{Josh Alman} {and}
        \bibinfo{person}{Virginia~Vassilevska Williams}.}
    \bibinfo{year}{2021}\natexlab{}.
    \newblock \showarticletitle{A Refined Laser Method and Faster Matrix
        Multiplication}. In \bibinfo{booktitle}{\emph{Proceedings of the 2021
            {ACM-SIAM} Symposium on Discrete Algorithms, {SODA}}},
    \bibfield{editor}{\bibinfo{person}{D{\'{a}}niel Marx}} (Ed.).
    \bibinfo{publisher}{{SIAM}}, \bibinfo{pages}{522--539}.
    \newblock
    
    
    \bibitem[\protect\citeauthoryear{B{\'{e}}res, P{\'{a}}lovics, Ol{\'{a}}h, and
        Bencz{\'{u}}r}{B{\'{e}}res et~al\mbox{.}}{2018}]%
    {katztg}
    \bibfield{author}{\bibinfo{person}{Ferenc B{\'{e}}res},
        \bibinfo{person}{R{\'{o}}bert P{\'{a}}lovics}, \bibinfo{person}{Anna
            Ol{\'{a}}h}, {and} \bibinfo{person}{Andr{\'{a}}s~A. Bencz{\'{u}}r}.}
    \bibinfo{year}{2018}\natexlab{}.
    \newblock \showarticletitle{Temporal walk based centrality metric for graph
        streams}.
    \newblock \bibinfo{journal}{\emph{Applied Network Science}}
    \bibinfo{volume}{3}, \bibinfo{number}{1} (\bibinfo{year}{2018}),
    \bibinfo{pages}{32:1--32:26}.
    \newblock
    
    
    \bibitem[\protect\citeauthoryear{Brandes}{Brandes}{2001}]%
    {brandes2001faster}
    \bibfield{author}{\bibinfo{person}{Ulrik Brandes}.}
    \bibinfo{year}{2001}\natexlab{}.
    \newblock \showarticletitle{A faster algorithm for betweenness centrality}.
    \newblock \bibinfo{journal}{\emph{The Journal of Mathematical Sociology}}
    \bibinfo{volume}{25}, \bibinfo{number}{2} (\bibinfo{year}{2001}),
    \bibinfo{pages}{163--177}.
    \newblock
    
    
    \bibitem[\protect\citeauthoryear{Brin and Page}{Brin and Page}{1998}]%
    {brin1998anatomy}
    \bibfield{author}{\bibinfo{person}{Sergey Brin} {and} \bibinfo{person}{Lawrence
            Page}.} \bibinfo{year}{1998}\natexlab{}.
    \newblock \showarticletitle{The anatomy of a large-scale hypertextual web
        search engine}.
    \newblock \bibinfo{journal}{\emph{Computer networks and ISDN systems}}
    \bibinfo{volume}{30}, \bibinfo{number}{1-7} (\bibinfo{year}{1998}),
    \bibinfo{pages}{107--117}.
    \newblock
    
    
    \bibitem[\protect\citeauthoryear{Bu{\ss}, Molter, Niedermeier, and
        Rymar}{Bu{\ss} et~al\mbox{.}}{2020}]%
    {BussMNR20}
    \bibfield{author}{\bibinfo{person}{Sebastian Bu{\ss}}, \bibinfo{person}{Hendrik
            Molter}, \bibinfo{person}{Rolf Niedermeier}, {and} \bibinfo{person}{Maciej
            Rymar}.} \bibinfo{year}{2020}\natexlab{}.
    \newblock \showarticletitle{Algorithmic Aspects of Temporal Betweenness}. In
    \bibinfo{booktitle}{\emph{{KDD} '20: The 26th {ACM} {SIGKDD} Conference on
            Knowledge Discovery and Data Mining}}. \bibinfo{publisher}{{ACM}},
    \bibinfo{pages}{2084--2092}.
    \newblock
    
    
    \bibitem[\protect\citeauthoryear{Coppersmith and Winograd}{Coppersmith and
        Winograd}{1987}]%
    {Coppersmith1987}
    \bibfield{author}{\bibinfo{person}{Don Coppersmith} {and}
        \bibinfo{person}{Shmuel Winograd}.} \bibinfo{year}{1987}\natexlab{}.
    \newblock \showarticletitle{Matrix multiplication via arithmetic progressions}.
    In \bibinfo{booktitle}{\emph{Proceedings of the nineteenth annual ACM
            symposium on Theory of computing}} \emph{(\bibinfo{series}{STOC '87})}.
    \bibinfo{publisher}{ACM}, \bibinfo{address}{New York, NY, USA},
    \bibinfo{pages}{1--6}.
    \newblock
    
    
    \bibitem[\protect\citeauthoryear{Crescenzi, Magnien, and Marino}{Crescenzi
        et~al\mbox{.}}{2020}]%
    {crescenzi2020finding}
    \bibfield{author}{\bibinfo{person}{Pierluigi Crescenzi},
        \bibinfo{person}{Cl{\'e}mence Magnien}, {and} \bibinfo{person}{Andrea
            Marino}.} \bibinfo{year}{2020}\natexlab{}.
    \newblock \showarticletitle{Finding top-k nodes for temporal closeness in large
        temporal graphs}.
    \newblock \bibinfo{journal}{\emph{Algorithms}} \bibinfo{volume}{13},
    \bibinfo{number}{9} (\bibinfo{year}{2020}), \bibinfo{pages}{211}.
    \newblock
    
    
    \bibitem[\protect\citeauthoryear{Das, Samanta, and Pal}{Das
        et~al\mbox{.}}{2018}]%
    {das2018study}
    \bibfield{author}{\bibinfo{person}{Kousik Das}, \bibinfo{person}{Sovan
            Samanta}, {and} \bibinfo{person}{Madhumangal Pal}.}
    \bibinfo{year}{2018}\natexlab{}.
    \newblock \showarticletitle{Study on centrality measures in social networks:
        {A} survey}.
    \newblock \bibinfo{journal}{\emph{Social Network Analysis and Mining}}
    \bibinfo{volume}{8}, \bibinfo{number}{1} (\bibinfo{year}{2018}),
    \bibinfo{pages}{13}.
    \newblock
    
    
    \bibitem[\protect\citeauthoryear{Freeman}{Freeman}{1977}]%
    {Freeman1977}
    \bibfield{author}{\bibinfo{person}{Linton~C. Freeman}.}
    \bibinfo{year}{1977}\natexlab{}.
    \newblock \showarticletitle{A Set of Measures of Centrality Based on
        Betweenness}.
    \newblock \bibinfo{journal}{\emph{Sociometry}}  \bibinfo{volume}{40}
    (\bibinfo{year}{1977}), \bibinfo{pages}{35--41}.
    \newblock
    
    
    \bibitem[\protect\citeauthoryear{G{\'o}mez}{G{\'o}mez}{2019}]%
    {gomez2019centrality}
    \bibfield{author}{\bibinfo{person}{Sergio G{\'o}mez}.}
    \bibinfo{year}{2019}\natexlab{}.
    \newblock \showarticletitle{Centrality in networks: finding the most important
        nodes}.
    \newblock In \bibinfo{booktitle}{\emph{Business and Consumer Analytics: New
            Ideas}}. \bibinfo{publisher}{Springer}, \bibinfo{pages}{401--433}.
    \newblock
    
    
    \bibitem[\protect\citeauthoryear{Grando, Noble, and Lamb}{Grando
        et~al\mbox{.}}{2016}]%
    {grando2016analysis}
    \bibfield{author}{\bibinfo{person}{Felipe Grando}, \bibinfo{person}{Diego
            Noble}, {and} \bibinfo{person}{Luis~C. Lamb}.}
    \bibinfo{year}{2016}\natexlab{}.
    \newblock \showarticletitle{An analysis of centrality measures for complex and
        social networks}. In \bibinfo{booktitle}{\emph{2016 IEEE Global
            Communications Conference (GLOBECOM)}}. IEEE, \bibinfo{pages}{1--6}.
    \newblock
    
    
    \bibitem[\protect\citeauthoryear{Grindrod, Parsons, Higham, and
        Estrada}{Grindrod et~al\mbox{.}}{2011}]%
    {grindrod2011communicability}
    \bibfield{author}{\bibinfo{person}{Peter Grindrod}, \bibinfo{person}{Mark~C.
            Parsons}, \bibinfo{person}{Desmond~J. Higham}, {and} \bibinfo{person}{Ernesto
            Estrada}.} \bibinfo{year}{2011}\natexlab{}.
    \newblock \showarticletitle{Communicability across evolving networks}.
    \newblock \bibinfo{journal}{\emph{Physical Review E}} \bibinfo{volume}{83},
    \bibinfo{number}{4} (\bibinfo{year}{2011}), \bibinfo{pages}{046120}.
    \newblock
    
    
    \bibitem[\protect\citeauthoryear{Haddadan, Menghini, Riondato, and
        Upfal}{Haddadan et~al\mbox{.}}{2021}]%
    {haddadan2021repbublik}
    \bibfield{author}{\bibinfo{person}{Shahrzad Haddadan},
        \bibinfo{person}{Cristina Menghini}, \bibinfo{person}{Matteo Riondato}, {and}
        \bibinfo{person}{Eli Upfal}.} \bibinfo{year}{2021}\natexlab{}.
    \newblock \showarticletitle{RePBubLik: Reducing Polarized Bubble Radius with
        Link Insertions}. In \bibinfo{booktitle}{\emph{Proceedings of the 14th ACM
            International Conference on Web Search and Data Mining}}.
    \bibinfo{pages}{139--147}.
    \newblock
    
    
    \bibitem[\protect\citeauthoryear{Harary and Norman}{Harary and Norman}{1960}]%
    {Harary1960}
    \bibfield{author}{\bibinfo{person}{Frank Harary} {and}
        \bibinfo{person}{Robert~Z. Norman}.} \bibinfo{year}{1960}\natexlab{}.
    \newblock \showarticletitle{Some properties of line digraphs}.
    \newblock \bibinfo{journal}{\emph{Rendiconti del Circolo Matematico di
            Palermo}} \bibinfo{volume}{9}, \bibinfo{number}{2} (\bibinfo{date}{May}
    \bibinfo{year}{1960}), \bibinfo{pages}{161--168}.
    \newblock
    
    
    \bibitem[\protect\citeauthoryear{Hogg and Lerman}{Hogg and Lerman}{2012}]%
    {hogg2012social}
    \bibfield{author}{\bibinfo{person}{Tad Hogg} {and} \bibinfo{person}{Kristina
            Lerman}.} \bibinfo{year}{2012}\natexlab{}.
    \newblock \showarticletitle{Social dynamics of digg}.
    \newblock \bibinfo{journal}{\emph{EPJ Data Science}} \bibinfo{volume}{1},
    \bibinfo{number}{1} (\bibinfo{year}{2012}), \bibinfo{pages}{5}.
    \newblock
    
    
    \bibitem[\protect\citeauthoryear{Holme}{Holme}{2015}]%
    {holme2015modern}
    \bibfield{author}{\bibinfo{person}{Petter Holme}.}
    \bibinfo{year}{2015}\natexlab{}.
    \newblock \showarticletitle{Modern temporal network theory: a colloquium}.
    \newblock \bibinfo{journal}{\emph{The European Physical Journal B}}
    \bibinfo{volume}{88}, \bibinfo{number}{9} (\bibinfo{year}{2015}),
    \bibinfo{pages}{234}.
    \newblock
    
    
    \bibitem[\protect\citeauthoryear{Isella, Stehl{\'e}, Barrat, Cattuto, Pinton,
        and Van~den Broeck}{Isella et~al\mbox{.}}{2011}]%
    {Isella2011}
    \bibfield{author}{\bibinfo{person}{Lorenzo Isella}, \bibinfo{person}{Juliette
            Stehl{\'e}}, \bibinfo{person}{Alain Barrat}, \bibinfo{person}{Ciro Cattuto},
        \bibinfo{person}{Jean-Fran{\c{c}}ois Pinton}, {and} \bibinfo{person}{Wouter
            Van~den Broeck}.} \bibinfo{year}{2011}\natexlab{}.
    \newblock \showarticletitle{What's in a crowd? {A}nalysis of face-to-face
        behavioral networks}.
    \newblock \bibinfo{journal}{\emph{Journal of Theoretical Biology}}
    \bibinfo{volume}{271}, \bibinfo{number}{1} (\bibinfo{year}{2011}),
    \bibinfo{pages}{166--180}.
    \newblock
    
    
    \bibitem[\protect\citeauthoryear{Katz}{Katz}{1953}]%
    {katz1953new}
    \bibfield{author}{\bibinfo{person}{Leo Katz}.} \bibinfo{year}{1953}\natexlab{}.
    \newblock \showarticletitle{A new status index derived from sociometric
        analysis}.
    \newblock \bibinfo{journal}{\emph{Psychometrika}} \bibinfo{volume}{18},
    \bibinfo{number}{1} (\bibinfo{year}{1953}), \bibinfo{pages}{39--43}.
    \newblock
    
    
    \bibitem[\protect\citeauthoryear{Kempe, Kleinberg, and Tardos}{Kempe
        et~al\mbox{.}}{2003}]%
    {Kempe2003}
    \bibfield{author}{\bibinfo{person}{David Kempe}, \bibinfo{person}{Jon~M.
            Kleinberg}, {and} \bibinfo{person}{{\'{E}}va Tardos}.}
    \bibinfo{year}{2003}\natexlab{}.
    \newblock \showarticletitle{Maximizing the spread of influence through a social
        network}. In \bibinfo{booktitle}{\emph{Proceedings of the Ninth {ACM}
            {SIGKDD} International Conference on Knowledge Discovery and Data Mining}},
    \bibfield{editor}{\bibinfo{person}{Lise Getoor}, \bibinfo{person}{Ted~E.
            Senator}, \bibinfo{person}{Pedro~M. Domingos}, {and}
        \bibinfo{person}{Christos Faloutsos}} (Eds.). \bibinfo{publisher}{{ACM}},
    \bibinfo{pages}{137--146}.
    \newblock
    
    
    \bibitem[\protect\citeauthoryear{Kendall}{Kendall}{1938}]%
    {kendall1938new}
    \bibfield{author}{\bibinfo{person}{Maurice~G. Kendall}.}
    \bibinfo{year}{1938}\natexlab{}.
    \newblock \showarticletitle{A new measure of rank correlation}.
    \newblock \bibinfo{journal}{\emph{Biometrika}} \bibinfo{volume}{30},
    \bibinfo{number}{1/2} (\bibinfo{year}{1938}), \bibinfo{pages}{81--93}.
    \newblock
    
    
    \bibitem[\protect\citeauthoryear{Khanna, Chaturvedi, and Soh}{Khanna
        et~al\mbox{.}}{2020}]%
    {khanna2020two}
    \bibfield{author}{\bibinfo{person}{Gaurav Khanna}, \bibinfo{person}{Sanjay~K.
            Chaturvedi}, {and} \bibinfo{person}{Sieteng Soh}.}
    \bibinfo{year}{2020}\natexlab{}.
    \newblock \showarticletitle{Two-terminal reliability analysis for time-evolving
        and predictable delay-tolerant networks}.
    \newblock \bibinfo{journal}{\emph{Recent Advances in Electrical \& Electronic
            Engineering (Formerly Recent Patents on Electrical \& Electronic
            Engineering)}} \bibinfo{volume}{13}, \bibinfo{number}{2}
    (\bibinfo{year}{2020}), \bibinfo{pages}{236--250}.
    \newblock
    
    
    \bibitem[\protect\citeauthoryear{Kim and Anderson}{Kim and Anderson}{2012}]%
    {kim2012temporal}
    \bibfield{author}{\bibinfo{person}{Hyoungshick Kim} {and} \bibinfo{person}{Ross
            Anderson}.} \bibinfo{year}{2012}\natexlab{}.
    \newblock \showarticletitle{Temporal node centrality in complex networks}.
    \newblock \bibinfo{journal}{\emph{Physical Review E}} \bibinfo{volume}{85},
    \bibinfo{number}{2} (\bibinfo{year}{2012}), \bibinfo{pages}{026107}.
    \newblock
    
    
    \bibitem[\protect\citeauthoryear{Klimt and Yang}{Klimt and Yang}{2004}]%
    {klimt2004enron}
    \bibfield{author}{\bibinfo{person}{Bryan Klimt} {and} \bibinfo{person}{Yiming
            Yang}.} \bibinfo{year}{2004}\natexlab{}.
    \newblock \showarticletitle{The enron corpus: A new dataset for email
        classification research}. In \bibinfo{booktitle}{\emph{European Conference on
            Machine Learning}}. Springer, \bibinfo{pages}{217--226}.
    \newblock
    
    
    \bibitem[\protect\citeauthoryear{Landherr, Friedl, and Heidemann}{Landherr
        et~al\mbox{.}}{2010}]%
    {landherr2010critical}
    \bibfield{author}{\bibinfo{person}{Andrea Landherr}, \bibinfo{person}{Bettina
            Friedl}, {and} \bibinfo{person}{Julia Heidemann}.}
    \bibinfo{year}{2010}\natexlab{}.
    \newblock \showarticletitle{A critical review of centrality measures in social
        networks}.
    \newblock \bibinfo{journal}{\emph{Business \& Information Systems Engineering}}
    \bibinfo{volume}{2}, \bibinfo{number}{6} (\bibinfo{year}{2010}),
    \bibinfo{pages}{371--385}.
    \newblock
    
    
    \bibitem[\protect\citeauthoryear{Latapy, Viard, and Magnien}{Latapy
        et~al\mbox{.}}{2018}]%
    {streamgraphs}
    \bibfield{author}{\bibinfo{person}{Matthieu Latapy}, \bibinfo{person}{Tiphaine
            Viard}, {and} \bibinfo{person}{Cl{\'{e}}mence Magnien}.}
    \bibinfo{year}{2018}\natexlab{}.
    \newblock \showarticletitle{Stream graphs and link streams for the modeling of
        interactions over time}.
    \newblock \bibinfo{journal}{\emph{Soc. Netw. Anal. Min.}} \bibinfo{volume}{8},
    \bibinfo{number}{1} (\bibinfo{year}{2018}), \bibinfo{pages}{61:1--61:29}.
    \newblock
    
    
    \bibitem[\protect\citeauthoryear{Liang and Modiano}{Liang and Modiano}{2016}]%
    {liang2016survivability}
    \bibfield{author}{\bibinfo{person}{Qingkai Liang} {and} \bibinfo{person}{Eytan
            Modiano}.} \bibinfo{year}{2016}\natexlab{}.
    \newblock \showarticletitle{Survivability in time-varying networks}.
    \newblock \bibinfo{journal}{\emph{IEEE Transactions on Mobile Computing}}
    \bibinfo{volume}{16}, \bibinfo{number}{9} (\bibinfo{year}{2016}),
    \bibinfo{pages}{2668--2681}.
    \newblock
    
    
    \bibitem[\protect\citeauthoryear{Marchiori and Latora}{Marchiori and
        Latora}{2000}]%
    {marchiori2000harmony}
    \bibfield{author}{\bibinfo{person}{Massimo Marchiori} {and}
        \bibinfo{person}{Vito Latora}.} \bibinfo{year}{2000}\natexlab{}.
    \newblock \showarticletitle{Harmony in the small-world}.
    \newblock \bibinfo{journal}{\emph{Physica A: Statistical Mechanics and its
            Applications}} \bibinfo{volume}{285}, \bibinfo{number}{3-4}
    (\bibinfo{year}{2000}), \bibinfo{pages}{539--546}.
    \newblock
    
    
    \bibitem[\protect\citeauthoryear{Mastrandrea, Fournet, and Barrat}{Mastrandrea
        et~al\mbox{.}}{2015}]%
    {mastrandrea2015contact}
    \bibfield{author}{\bibinfo{person}{Rossana Mastrandrea}, \bibinfo{person}{Julie
            Fournet}, {and} \bibinfo{person}{Alain Barrat}.}
    \bibinfo{year}{2015}\natexlab{}.
    \newblock \showarticletitle{Contact patterns in a high school: a comparison
        between data collected using wearable sensors, contact diaries and friendship
        surveys}.
    \newblock \bibinfo{journal}{\emph{PloS one}} \bibinfo{volume}{10},
    \bibinfo{number}{9} (\bibinfo{year}{2015}), \bibinfo{pages}{e0136497}.
    \newblock
    
    
    \bibitem[\protect\citeauthoryear{Mislove}{Mislove}{2009}]%
    {mislove2009online}
    \bibfield{author}{\bibinfo{person}{Alan~E. Mislove}.}
    \bibinfo{year}{2009}\natexlab{}.
    \newblock \emph{\bibinfo{title}{Online social networks: measurement, analysis,
            and applications to distributed information systems}}.
    \newblock \bibinfo{thesistype}{Ph.\,D. Dissertation}. \bibinfo{school}{Rice
        University}.
    \newblock
    
    
    \bibitem[\protect\citeauthoryear{Mislove, Marcon, Gummadi, Druschel, and
        Bhattacharjee}{Mislove et~al\mbox{.}}{2007}]%
    {MisloveMGDB07}
    \bibfield{author}{\bibinfo{person}{Alan~E. Mislove},
        \bibinfo{person}{Massimiliano Marcon}, \bibinfo{person}{P.~Krishna Gummadi},
        \bibinfo{person}{Peter Druschel}, {and} \bibinfo{person}{Bobby
            Bhattacharjee}.} \bibinfo{year}{2007}\natexlab{}.
    \newblock \showarticletitle{Measurement and analysis of online social
        networks}. In \bibinfo{booktitle}{\emph{Proceedings of the 7th {ACM}
            {SIGCOMM} Internet Measurement Conference, {IMC}}}.
    \bibinfo{publisher}{{ACM}}, \bibinfo{pages}{29--42}.
    \newblock
    
    
    \bibitem[\protect\citeauthoryear{Mutzel and Oettershagen}{Mutzel and
        Oettershagen}{2019}]%
    {mutzel2019enumeration}
    \bibfield{author}{\bibinfo{person}{Petra Mutzel} {and} \bibinfo{person}{Lutz
            Oettershagen}.} \bibinfo{year}{2019}\natexlab{}.
    \newblock \showarticletitle{On the enumeration of bicriteria temporal paths}.
    In \bibinfo{booktitle}{\emph{International Conference on Theory and
            Applications of Models of Computation}}. Springer, \bibinfo{pages}{518--535}.
    \newblock
    
    
    \bibitem[\protect\citeauthoryear{Newman}{Newman}{2005}]%
    {Newman2005}
    \bibfield{author}{\bibinfo{person}{Mark E.~J. Newman}.}
    \bibinfo{year}{2005}\natexlab{}.
    \newblock \showarticletitle{A measure of betweenness centrality based on random
        walks}.
    \newblock \bibinfo{journal}{\emph{Soc. Networks}} \bibinfo{volume}{27},
    \bibinfo{number}{1} (\bibinfo{year}{2005}), \bibinfo{pages}{39--54}.
    \newblock
    
    
    \bibitem[\protect\citeauthoryear{Newman}{Newman}{2010}]%
    {Newman2010}
    \bibfield{author}{\bibinfo{person}{Mark E.~J. Newman}.}
    \bibinfo{year}{2010}\natexlab{}.
    \newblock \bibinfo{booktitle}{\emph{Networks: An Introduction}}.
    \newblock \bibinfo{publisher}{Oxford University Press}.
    \newblock
    \showISBNx{978-0-19920665-0}
    
    
    \bibitem[\protect\citeauthoryear{Nguyen, Lee, Rossi, Ahmed, Koh, and
        Kim}{Nguyen et~al\mbox{.}}{2018}]%
    {nguyen2018continuous}
    \bibfield{author}{\bibinfo{person}{Giang~Hoang Nguyen},
        \bibinfo{person}{John~Boaz Lee}, \bibinfo{person}{Ryan~A. Rossi},
        \bibinfo{person}{Nesreen~K. Ahmed}, \bibinfo{person}{Eunyee Koh}, {and}
        \bibinfo{person}{Sungchul Kim}.} \bibinfo{year}{2018}\natexlab{}.
    \newblock \showarticletitle{Continuous-time dynamic network embeddings}. In
    \bibinfo{booktitle}{\emph{Companion Proceedings of the The Web Conference
            2018}}. \bibinfo{pages}{969--976}.
    \newblock
    
    
    \bibitem[\protect\citeauthoryear{Nicosia, Tang, Mascolo, Musolesi, Russo, and
        Latora}{Nicosia et~al\mbox{.}}{2013}]%
    {nicosia2013graph}
    \bibfield{author}{\bibinfo{person}{Vincenzo Nicosia}, \bibinfo{person}{John
            Tang}, \bibinfo{person}{Cecilia Mascolo}, \bibinfo{person}{Mirco Musolesi},
        \bibinfo{person}{Giovanni Russo}, {and} \bibinfo{person}{Vito Latora}.}
    \bibinfo{year}{2013}\natexlab{}.
    \newblock \showarticletitle{Graph metrics for temporal networks}.
    \newblock In \bibinfo{booktitle}{\emph{Temporal {N}etworks}}.
    \bibinfo{publisher}{Springer}, \bibinfo{pages}{15--40}.
    \newblock
    
    
    \bibitem[\protect\citeauthoryear{Oettershagen, Kriege, Morris, and
        Mutzel}{Oettershagen et~al\mbox{.}}{2020}]%
    {oettershagen2020temporal}
    \bibfield{author}{\bibinfo{person}{Lutz Oettershagen}, \bibinfo{person}{Nils~M.
            Kriege}, \bibinfo{person}{Christopher Morris}, {and} \bibinfo{person}{Petra
            Mutzel}.} \bibinfo{year}{2020}\natexlab{}.
    \newblock \showarticletitle{Temporal Graph Kernels for Classifying
        Dissemination Processes}. In \bibinfo{booktitle}{\emph{Proceedings of the
            2020 SIAM International Conference on Data Mining}}. SIAM,
    \bibinfo{pages}{496--504}.
    \newblock
    
    
    \bibitem[\protect\citeauthoryear{Oettershagen and Mutzel}{Oettershagen and
        Mutzel}{2020}]%
    {oettershagen2020efficient}
    \bibfield{author}{\bibinfo{person}{Lutz Oettershagen} {and}
        \bibinfo{person}{Petra Mutzel}.} \bibinfo{year}{2020}\natexlab{}.
    \newblock \showarticletitle{Efficient Top-k Temporal Closeness Calculation in
        Temporal Networks}. In \bibinfo{booktitle}{\emph{2020 IEEE International
            Conference on Data Mining (ICDM)}}. IEEE, \bibinfo{pages}{402--411}.
    \newblock
    
    
    \bibitem[\protect\citeauthoryear{Oettershagen and Mutzel}{Oettershagen and
        Mutzel}{2022}]%
    {oettershagen2022computing}
    \bibfield{author}{\bibinfo{person}{Lutz Oettershagen} {and}
        \bibinfo{person}{Petra Mutzel}.} \bibinfo{year}{2022}\natexlab{}.
    \newblock \showarticletitle{Computing top-k temporal closeness in temporal
        networks}.
    \newblock \bibinfo{journal}{\emph{Knowledge and Information Systems}}
    (\bibinfo{year}{2022}), \bibinfo{pages}{1--29}.
    \newblock
    
    
    \bibitem[\protect\citeauthoryear{Opsahl and Panzarasa}{Opsahl and
        Panzarasa}{2009}]%
    {opsahl2009clustering}
    \bibfield{author}{\bibinfo{person}{Tore Opsahl} {and} \bibinfo{person}{Pietro
            Panzarasa}.} \bibinfo{year}{2009}\natexlab{}.
    \newblock \showarticletitle{Clustering in weighted networks}.
    \newblock \bibinfo{journal}{\emph{Social networks}} \bibinfo{volume}{31},
    \bibinfo{number}{2} (\bibinfo{year}{2009}), \bibinfo{pages}{155--163}.
    \newblock
    
    
    \bibitem[\protect\citeauthoryear{Page, Brin, Motwani, and Winograd}{Page
        et~al\mbox{.}}{1999}]%
    {Page1999}
    \bibfield{author}{\bibinfo{person}{Lawrence Page}, \bibinfo{person}{Sergey
            Brin}, \bibinfo{person}{Rajeev Motwani}, {and} \bibinfo{person}{Terry
            Winograd}.} \bibinfo{year}{1999}\natexlab{}.
    \newblock \bibinfo{booktitle}{\emph{The PageRank Citation Ranking: Bringing
            Order to the Web.}}
    \newblock \bibinfo{type}{Technical Report} 1999-66.
    \bibinfo{institution}{Stanford InfoLab}.
    \newblock
    \newblock
    \shownote{Previous number = SIDL-WP-1999-0120}.
    
    
    \bibitem[\protect\citeauthoryear{Panzarasa, Opsahl, and Carley}{Panzarasa
        et~al\mbox{.}}{2009}]%
    {panzarasa2009patterns}
    \bibfield{author}{\bibinfo{person}{Pietro Panzarasa}, \bibinfo{person}{Tore
            Opsahl}, {and} \bibinfo{person}{Kathleen~M. Carley}.}
    \bibinfo{year}{2009}\natexlab{}.
    \newblock \showarticletitle{Patterns and dynamics of users' behavior and
        interaction: Network analysis of an online community}.
    \newblock \bibinfo{journal}{\emph{Journal of the American Society for
            Information Science and Technology}} \bibinfo{volume}{60},
    \bibinfo{number}{5} (\bibinfo{year}{2009}), \bibinfo{pages}{911--932}.
    \newblock
    
    
    \bibitem[\protect\citeauthoryear{Paranjape, Benson, and Leskovec}{Paranjape
        et~al\mbox{.}}{2017}]%
    {paranjape2017motifs}
    \bibfield{author}{\bibinfo{person}{Ashwin Paranjape},
        \bibinfo{person}{Austin~R. Benson}, {and} \bibinfo{person}{Jure Leskovec}.}
    \bibinfo{year}{2017}\natexlab{}.
    \newblock \showarticletitle{Motifs in temporal networks}. In
    \bibinfo{booktitle}{\emph{Proceedings of the Tenth ACM International
            Conference on Web Search and Data Mining}}. \bibinfo{pages}{601--610}.
    \newblock
    
    
    \bibitem[\protect\citeauthoryear{Richardson, Agrawal, and Domingos}{Richardson
        et~al\mbox{.}}{2003}]%
    {richardson2003trust}
    \bibfield{author}{\bibinfo{person}{Matthew Richardson}, \bibinfo{person}{Rakesh
            Agrawal}, {and} \bibinfo{person}{Pedro Domingos}.}
    \bibinfo{year}{2003}\natexlab{}.
    \newblock \showarticletitle{Trust management for the semantic web}.
    \newblock In \bibinfo{booktitle}{\emph{The Semantic Web-ISWC 2003}}.
    \bibinfo{publisher}{Springer}, \bibinfo{pages}{351--368}.
    \newblock
    
    
    \bibitem[\protect\citeauthoryear{Rodrigues}{Rodrigues}{2019}]%
    {rodrigues2019network}
    \bibfield{author}{\bibinfo{person}{Francisco~Aparecido Rodrigues}.}
    \bibinfo{year}{2019}\natexlab{}.
    \newblock \showarticletitle{Network centrality: {A}n introduction}.
    \newblock In \bibinfo{booktitle}{\emph{A Mathematical Modeling Approach from
            Nonlinear Dynamics to Complex Systems}}. \bibinfo{publisher}{Springer},
    \bibinfo{pages}{177--196}.
    \newblock
    
    
    \bibitem[\protect\citeauthoryear{Ronqui and Travieso}{Ronqui and
        Travieso}{2015}]%
    {ronqui2015analyzing}
    \bibfield{author}{\bibinfo{person}{Jos{\'e} Ricardo~Furlan Ronqui} {and}
        \bibinfo{person}{Gonzalo Travieso}.} \bibinfo{year}{2015}\natexlab{}.
    \newblock \showarticletitle{Analyzing complex networks through correlations in
        centrality measurements}.
    \newblock \bibinfo{journal}{\emph{Journal of Statistical Mechanics: Theory and
            Experiment}} \bibinfo{volume}{2015}, \bibinfo{number}{5}
    (\bibinfo{year}{2015}), \bibinfo{pages}{P05030}.
    \newblock
    
    
    \bibitem[\protect\citeauthoryear{Rozenshtein and Gionis}{Rozenshtein and
        Gionis}{2016}]%
    {DBLP:conf/pkdd/RozenshteinG16}
    \bibfield{author}{\bibinfo{person}{Polina Rozenshtein} {and}
        \bibinfo{person}{Aristides Gionis}.} \bibinfo{year}{2016}\natexlab{}.
    \newblock \showarticletitle{Temporal PageRank}. In
    \bibinfo{booktitle}{\emph{Machine Learning and Knowledge Discovery in
            Databases - European Conference, {ECML} {PKDD} 2016}},
    Vol.~\bibinfo{volume}{9852}. \bibinfo{publisher}{Springer},
    \bibinfo{pages}{674--689}.
    \newblock
    
    
    \bibitem[\protect\citeauthoryear{Santoro, Quattrociocchi, Flocchini, Casteigts,
        and Amblard}{Santoro et~al\mbox{.}}{2011}]%
    {santoro2011time}
    \bibfield{author}{\bibinfo{person}{Nicola Santoro}, \bibinfo{person}{Walter
            Quattrociocchi}, \bibinfo{person}{Paola Flocchini}, \bibinfo{person}{Arnaud
            Casteigts}, {and} \bibinfo{person}{Frederic Amblard}.}
    \bibinfo{year}{2011}\natexlab{}.
    \newblock \showarticletitle{Time-varying graphs and social network analysis:
        Temporal indicators and metrics}.
    \newblock \bibinfo{journal}{\emph{arXiv preprint arXiv:1102.0629}}
    (\bibinfo{year}{2011}).
    \newblock
    
    
    \bibitem[\protect\citeauthoryear{Saxena and Iyengar}{Saxena and
        Iyengar}{2020}]%
    {Saxena2020}
    \bibfield{author}{\bibinfo{person}{Akrati Saxena} {and}
        \bibinfo{person}{Sudarshan Iyengar}.} \bibinfo{year}{2020}\natexlab{}.
    \newblock \showarticletitle{Centrality Measures in Complex Networks: {A}
        Survey}.
    \newblock \bibinfo{journal}{\emph{CoRR}}  \bibinfo{volume}{abs/2011.07190}
    (\bibinfo{year}{2020}).
    \newblock
    \showeprint[arxiv]{2011.07190}
    \urldef\tempurl%
    \url{https://arxiv.org/abs/2011.07190}
    \showURL{%
        \tempurl}
    
    
    \bibitem[\protect\citeauthoryear{Sun, Kunegis, and Staab}{Sun
        et~al\mbox{.}}{2016}]%
    {SunKS16}
    \bibfield{author}{\bibinfo{person}{Jun Sun},
        \bibinfo{person}{J{\'{e}}r{\^{o}}me Kunegis}, {and} \bibinfo{person}{Steffen
            Staab}.} \bibinfo{year}{2016}\natexlab{}.
    \newblock \showarticletitle{Predicting User Roles in Social Networks Using
        Transfer Learning with Feature Transformation}. In
    \bibinfo{booktitle}{\emph{{IEEE} International Conference on Data Mining
            Workshops, {ICDM} Workshops}}. \bibinfo{publisher}{{IEEE} Computer Society},
    \bibinfo{pages}{128--135}.
    \newblock
    
    
    \bibitem[\protect\citeauthoryear{Tang, Leontiadis, Scellato, Nicosia, Mascolo,
        Musolesi, and Latora}{Tang et~al\mbox{.}}{2013}]%
    {tang2013applications}
    \bibfield{author}{\bibinfo{person}{John Tang}, \bibinfo{person}{Ilias
            Leontiadis}, \bibinfo{person}{Salvatore Scellato}, \bibinfo{person}{Vincenzo
            Nicosia}, \bibinfo{person}{Cecilia Mascolo}, \bibinfo{person}{Mirco
            Musolesi}, {and} \bibinfo{person}{Vito Latora}.}
    \bibinfo{year}{2013}\natexlab{}.
    \newblock \showarticletitle{Applications of temporal graph metrics to
        real-world networks}.
    \newblock In \bibinfo{booktitle}{\emph{Temporal Networks}}.
    \bibinfo{publisher}{Springer}, \bibinfo{pages}{135--159}.
    \newblock
    
    
    \bibitem[\protect\citeauthoryear{Tang, Musolesi, Mascolo, Latora, and
        Nicosia}{Tang et~al\mbox{.}}{2010}]%
    {tang2010analysing}
    \bibfield{author}{\bibinfo{person}{John Tang}, \bibinfo{person}{Mirco
            Musolesi}, \bibinfo{person}{Cecilia Mascolo}, \bibinfo{person}{Vito Latora},
        {and} \bibinfo{person}{Vincenzo Nicosia}.} \bibinfo{year}{2010}\natexlab{}.
    \newblock \showarticletitle{Analysing information flows and key mediators
        through temporal centrality metrics}. In \bibinfo{booktitle}{\emph{Proc.\ 3rd
            Workshop on Social Network Systems}}. \bibinfo{pages}{1--6}.
    \newblock
    
    
    \bibitem[\protect\citeauthoryear{Tsalouchidou, Baeza-Yates, Bonchi, Liao, and
        Sellis}{Tsalouchidou et~al\mbox{.}}{2019}]%
    {tsalouchidou2019temporal}
    \bibfield{author}{\bibinfo{person}{Ioanna Tsalouchidou},
        \bibinfo{person}{Ricardo Baeza-Yates}, \bibinfo{person}{Francesco Bonchi},
        \bibinfo{person}{Kewen Liao}, {and} \bibinfo{person}{Timos Sellis}.}
    \bibinfo{year}{2019}\natexlab{}.
    \newblock \showarticletitle{Temporal betweenness centrality in dynamic graphs}.
    \newblock \bibinfo{journal}{\emph{International Journal of Data Science and
            Analytics}} (\bibinfo{year}{2019}), \bibinfo{pages}{1--16}.
    \newblock
    
    
    \bibitem[\protect\citeauthoryear{Vanhems, Barrat, Cattuto, Pinton, Khanafer,
        R{\'e}gis, Kim, Comte, and Voirin}{Vanhems et~al\mbox{.}}{2013}]%
    {vanhems2013estimating}
    \bibfield{author}{\bibinfo{person}{Philippe Vanhems}, \bibinfo{person}{Alain
            Barrat}, \bibinfo{person}{Ciro Cattuto}, \bibinfo{person}{Jean-Fran{\c{c}}ois
            Pinton}, \bibinfo{person}{Nagham Khanafer}, \bibinfo{person}{Corinne
            R{\'e}gis}, \bibinfo{person}{Byeul-a Kim}, \bibinfo{person}{Brigitte Comte},
        {and} \bibinfo{person}{Nicolas Voirin}.} \bibinfo{year}{2013}\natexlab{}.
    \newblock \showarticletitle{Estimating potential infection transmission routes
        in hospital wards using wearable proximity sensors}.
    \newblock \bibinfo{journal}{\emph{PloS one}} \bibinfo{volume}{8},
    \bibinfo{number}{9} (\bibinfo{year}{2013}), \bibinfo{pages}{e73970}.
    \newblock
    
    
    \bibitem[\protect\citeauthoryear{Viswanath, Mislove, Cha, and
        Gummadi}{Viswanath et~al\mbox{.}}{2009}]%
    {viswanath2009evolution}
    \bibfield{author}{\bibinfo{person}{Bimal Viswanath}, \bibinfo{person}{Alan
            Mislove}, \bibinfo{person}{Meeyoung Cha}, {and} \bibinfo{person}{Krishna~P.
            Gummadi}.} \bibinfo{year}{2009}\natexlab{}.
    \newblock \showarticletitle{On the evolution of user interaction in facebook}.
    In \bibinfo{booktitle}{\emph{Proceedings of the 2nd ACM workshop on Online
            social networks}}. \bibinfo{pages}{37--42}.
    \newblock
    
    
    \bibitem[\protect\citeauthoryear{Wetzker, Zimmermann, and Bauckhage}{Wetzker
        et~al\mbox{.}}{2008}]%
    {wetzker2008analyzing}
    \bibfield{author}{\bibinfo{person}{Robert Wetzker}, \bibinfo{person}{Carsten
            Zimmermann}, {and} \bibinfo{person}{Christian Bauckhage}.}
    \bibinfo{year}{2008}\natexlab{}.
    \newblock \showarticletitle{Analyzing social bookmarking systems: A del. icio.
        us cookbook}. In \bibinfo{booktitle}{\emph{Proceedings of the ECAI 2008
            Mining Social Data Workshop}}. \bibinfo{pages}{26--30}.
    \newblock
    
    
    \bibitem[\protect\citeauthoryear{White and Smyth}{White and Smyth}{2003}]%
    {white2003algorithms}
    \bibfield{author}{\bibinfo{person}{Scott White} {and} \bibinfo{person}{Padhraic
            Smyth}.} \bibinfo{year}{2003}\natexlab{}.
    \newblock \showarticletitle{Algorithms for estimating relative importance in
        networks}. In \bibinfo{booktitle}{\emph{Proceedings of the ninth ACM SIGKDD
            international conference on Knowledge discovery and data mining}}.
    \bibinfo{pages}{266--275}.
    \newblock
    
    
    \bibitem[\protect\citeauthoryear{Wu, Cheng, Huang, Ke, Lu, and Xu}{Wu
        et~al\mbox{.}}{2014}]%
    {wu2014path}
    \bibfield{author}{\bibinfo{person}{Huanhuan Wu}, \bibinfo{person}{James Cheng},
        \bibinfo{person}{Silu Huang}, \bibinfo{person}{Yiping Ke},
        \bibinfo{person}{Yi Lu}, {and} \bibinfo{person}{Yanyan Xu}.}
    \bibinfo{year}{2014}\natexlab{}.
    \newblock \showarticletitle{Path problems in temporal graphs}.
    \newblock \bibinfo{journal}{\emph{Proc.\ VLDB Endowment}} \bibinfo{volume}{7},
    \bibinfo{number}{9} (\bibinfo{year}{2014}), \bibinfo{pages}{721--732}.
    \newblock
    
    
    \bibitem[\protect\citeauthoryear{Wu, Fu, Zhang, Long, Meng, Wang, and Chen}{Wu
        et~al\mbox{.}}{2020}]%
    {Wu2020}
    \bibfield{author}{\bibinfo{person}{Xudong Wu}, \bibinfo{person}{Luoyi Fu},
        \bibinfo{person}{Zixin Zhang}, \bibinfo{person}{Huan Long},
        \bibinfo{person}{Jingfan Meng}, \bibinfo{person}{Xinbing Wang}, {and}
        \bibinfo{person}{Guihai Chen}.} \bibinfo{year}{2020}\natexlab{}.
    \newblock \showarticletitle{Evolving Influence Maximization in Evolving
        Networks}.
    \newblock \bibinfo{journal}{\emph{ACM Trans. Internet Technol.}}
    \bibinfo{volume}{20}, \bibinfo{number}{4}, Article \bibinfo{articleno}{40}
    (\bibinfo{date}{Oct.} \bibinfo{year}{2020}).
    \newblock
    \showISSN{1533-5399}
    
    
\end{thebibliography}
\end{document}